\documentclass[openacc]{rstransa}

\begin{document}

\title{The cosmological constant and scale hierarchies with emergent gauge symmetries
}

\author{
Steven D. Bass$^{1,2}$
}

\address{$^{1}$Kitzb\"uhel Centre for Physics, Kitzb\"uhel Austria \\
$^2$
Marian Smoluchowski Institute of Physics,  Jagiellonian University, Krak\'ow, Poland \\
}

\subject{Particle physics, gauge symmetries, cosmology}


\corres{Steven D. Bass\\
\email{Steven.Bass@cern.ch}}

\begin{abstract}
Motivated by the stability of the electroweak Higgs 
vacuum
we consider the possibility that the Standard Model might work up to 
large scales between about $10^{10}$ GeV and close to the Planck scale. 
A plausible scenario is an emergent Standard Model 
with gauge symmetries originating in some topological like phase transition deep in the ultraviolet. 
In this case 
the cosmological constant scale and neutrino masses should be of similar size, suppressed by factor of the large scale of emergence.
The key physics involves a subtle interplay of 
Poincar\'e invariance, mass generation and renormalisation group invariance. 
The Higgs mass would be environmentally selected in connection with vacuum stability.
Consequences for dark matter scenarios are discussed. 

\end{abstract}


\begin{fmtext}

\end{fmtext}
%
%
\maketitle

\section{Introduction}
\label{sec:Introduction}

Our observations in 
particle physics and gravitation are described by the Standard Model 
and General Relativity. 
The Standard Model \cite{Pokorski:1987ed,Taylor:1976ru}
works very well so far 
in all experiments 
from LHC collider physics
\cite{Altarelli:2013tya,Bass:2021acr,ATLAS:2022vkf,CMS:2022dwd}
to precision measurements of, e.g., 
the fine structure constant \cite{Fan:2022eto}
and the electron electric dipole moment
\cite{Roussy:2022cmp}
with so far no evidence for new particles or interactions in the range of present experiments. 
The Standard Model is  built on the local 
gauge symmetries of 
electroweak interactions and quantum chromodynamics, QCD,  with the gauge groups 
U(1)$_Y$, SU(2)$_L$ and SU(3)$_c$. 
Beyond the spin 
$\frac{1}{2}$ leptons and quarks plus spin-one gauge bosons, 
the Higgs boson with spin zero 
and mass 125 GeV discovered at CERN behaves very Standard Model like and completes the particle spectrum of the Standard Model \cite{Bass:2021acr,Jakobs:2023}.

General Relativity
describes gravitation wherever it has been tested 
with recent highlights including 
the discovery of gravitational waves 
\cite{LIGOScientific:2016aoc}
and 
black hole imaging
\cite{EventHorizonTelescope:2019dse}. 
In the laboratory Newton's law works 
down to at least 52 $\mu$m~\cite{Lee:2020zjt}.
General Relativity is  also a gauge theory, under local transformations of the co-ordinate system~\cite{Kibble:1961ba,Sciama:1964wt}.

Yet we know this is not the complete story.
Some extra new physics is needed to understand 
issues with neutrinos 
(their masses, whether they might be Dirac or Majorana particles, possible CP phases), baryogenesis (why there is more matter than antimatter), 
the absence of strong CP violation, fermion families, 
dark energy (the accelerating expansion of the Universe) and dark matter as well as primordial inflation.

How high in energy might the Standard Model 
work before one runs into new physics?
The energy scale of any new interactions
is so far not known.
Where do the gauge symmetries come from at a deeper level? 
What about the dynamics behind 
electroweak symmetry breaking in the Standard Model? 
How should we understand the scale hierarchies of particle physics with the cosmological constant scale 0.002 eV very much less than the electroweak scale 246 GeV 
which is itself very much less than the Planck scale 
$M_{\rm Pl} = 1.2 \times 10^{19}$ GeV
where quantum gravity effects might apply?

Today a vigorous programme of experiments and theory is probing the high energy and precision frontiers
together with 
cosmology 
looking for cracks in our 
description of Nature provided by the 
Standard Model and General Relativity. 
Ideas for new physics are discussed in the parallel 
theoretical contributions to this volume  
\cite{Pokorski:2023,Dvali:2023,Heisenberg:2023} 
with experiments discussed in 
Refs.~\cite{Jakobs:2023,Baudis:2023,Ackermann:2023,Krizan:2023,Malbrunot:2023,Schieck:2023}.

While so far not revealing evidence for new particles and interactions, 
LHC data do suggest an intriguing result with stability of the Standard Model vacuum.
If we assume no coupling to new particles at higher energies 
and extrapolate the Standard Model into the deep ultraviolet using renormalisation group, RG, evolution, 
then the vacuum remains stable up to very high scales - at least $10^{10}$ GeV and perhaps up to the Planck scale with vacuum stability very sensitive to the exact values of Standard Model parameters 
\cite{Jegerlehner:2013cta,Bednyakov:2015sca}. 
Might the Standard Model  be more special than previously assumed?
Here we consider this possibility.
In particular, we discuss the idea that the 
particles and interactions of the 
Standard Model might be emergent 
below the scale characterising
some topological like phase transition deep in the ultraviolet 
\cite{Jegerlehner:2013cta,Bjorken:2001pe,Bass:2021wxv,Jegerlehner:2018zxm,Jegerlehner:2021vqz}.
In this scenario the Higgs mass is environmentally selected, connected to the stability of the vacuum,
with subtle connection between the 
infrared world of our experiments and the physics in the extreme ultraviolet. 
Interestingly, the cosmological constant  scale comes out similar to the size of tiny 
Majorana neutrino masses
\cite{Bass:2020egf,Bass:2020nrg}.
New global symmetry breaking interactions can occur in higher dimensional operators, suppressed by powers of the large scale of emergence - in this paper taken about $10^{16}$ GeV - which may have been active in the very early Universe and perhaps relevant to cosmology.
In the absence of larger multiplets at mass dimension four, 
dark matter might involve either primordial black
holes or new physics coupled through higher dimensional operators with small 
non-gravitational
coupling to 
normal baryonic matter.

The plan of this paper is as follows. 
In Section \ref{sec:VS} we briefly describe ideas about
the vacuum stability of the Standard Model.
Section \ref{sec:EGS} describes the phenomenology of a possible emergent Standard Model including the cosmological constant and dark matter scenarios. 
Sections \ref{sec:CC} and \ref{sec:Scales} are of more theoretical interest. 
Section \ref{sec:CC} discusses separate contributions to the cosmological constant and how they might fit together to give a small value of the cosmological constant. 
Related issues with hierarchies of scales in particle physics including 
the Higgs mass 
are described in Section \ref{sec:Scales}. 
Finally, in Section \ref{sec:conclusions} we summarise and conclude.

\section{Vacuum stability}
\label{sec:VS}

If the Standard Model is evolved using RG up towards the Planck scale 
assuming no coupling to extra particles, then one finds that the parameters measured in experiments  and its 
ultraviolet behaviour are strongly correlated.

The Standard Model couplings and particle masses 
are related.
For the W and Z gauge bosons
\begin{equation}
m_{\rm W}^2 = \frac{1}{4} g^2 v^2 , 
\ \ \ 
m_{\rm Z}^2 = \frac{1}{4} (g^2 + g'^2 )v^2 
\label{eq:2a}
\end{equation}
where 
$g$ and $g'$ are the SU(2) and U(1) electroweak couplings; 
$v$ is the Higgs vacuum expectation value 246 GeV.
The charged fermion masses are 
\begin{equation}
m_f = y_f \frac{v}{\sqrt{2}}  \ \ \ \ \ 
(f = {\rm quarks \ and \ charged \ leptons})
\label{eq:2b}
\end{equation}
where $y_f$ are the Yukawa couplings.
The Higgs mass is
\begin{equation}
m_h^2 = 2 \lambda v^2
\label{eq:2c}
\end{equation}
with $\lambda$ the Higgs self-coupling.

The masses of the W and Z and also the quarks and charged leptons come from the Brout-Englert-Higgs, BEH,  mechanism~\cite{Higgs:1964ia,Higgs:1964pj,Higgs:1966ev,Englert:1964et,Veltman:1997nm,Kibble:2014gug}.
These particles come with gauge invariant coupling to a 
complex Higgs doublet field $\phi$ 
with potential 
$
    V(\phi) = \frac{1}{2} m^2 \phi^*\phi
    + \frac{1}{4} \lambda (\phi^*\phi)^2
$ 
and $m^2 < 0$.
The Higgs field acquires a vacuum expectation value 
$|\phi| = v = \sqrt{-m^2/ \lambda} = 
m_h/\sqrt{2 \lambda}$ 
at the minimum of the potential
implying spontaneous symmetry breaking phenomena. 
Working in unitary gauge, 
of the four Higgs doublet components 
three Goldstone states are absorbed as the longitudinal  components of the W and Z with the remaining scalar degree of freedom 
being the massive Higgs boson. 
Here spontaneous symmetry breaking is defined relative to the choice of gauge, 
e.g. the unitary gauge, 
with all gauge choices being physically equivalent \cite{Kibble:2014gug}.

The Standard Model relations 
Eqs.~(\ref{eq:2a},\ref{eq:2b})
have been tested 
to good precision 
about 10\%
for the W and Z gauge bosons, the heavy top and bottom quarks, 
and 
the $\tau$ and $\mu$ charged leptons.  
The measured coupling strengths scale as a function of the particle masses just 
as predicted by the Standard Model  
-- see 
\cite{Bass:2021acr,Jakobs:2023}.
For RG evolution one also
needs input on 
the Higgs self-coupling $\lambda$. 
In the absence of direct measurement,  
here we have to assume 
the Standard Model relation connecting 
$\lambda$ to the Higgs mass, 
Eq.~(\ref{eq:2c}).
Taking $v=246$ GeV and $m_h=125$ GeV 
gives 
$\lambda = m_h^2/2 v^2 
\approx 0.13$ 
as input 
at LHC laboratory scales. 
Measurement of $\lambda$ awaits future 
collider experiments 
with
precision about 50\% 
expected from the high luminosity upgrade of the LHC and 
5\% precision  
requiring a new 100 TeV 
higher energy collider \cite{Jakobs:2023}.

\begin{figure}[t!]  
\centerline
{\includegraphics[width=0.53\textwidth]
{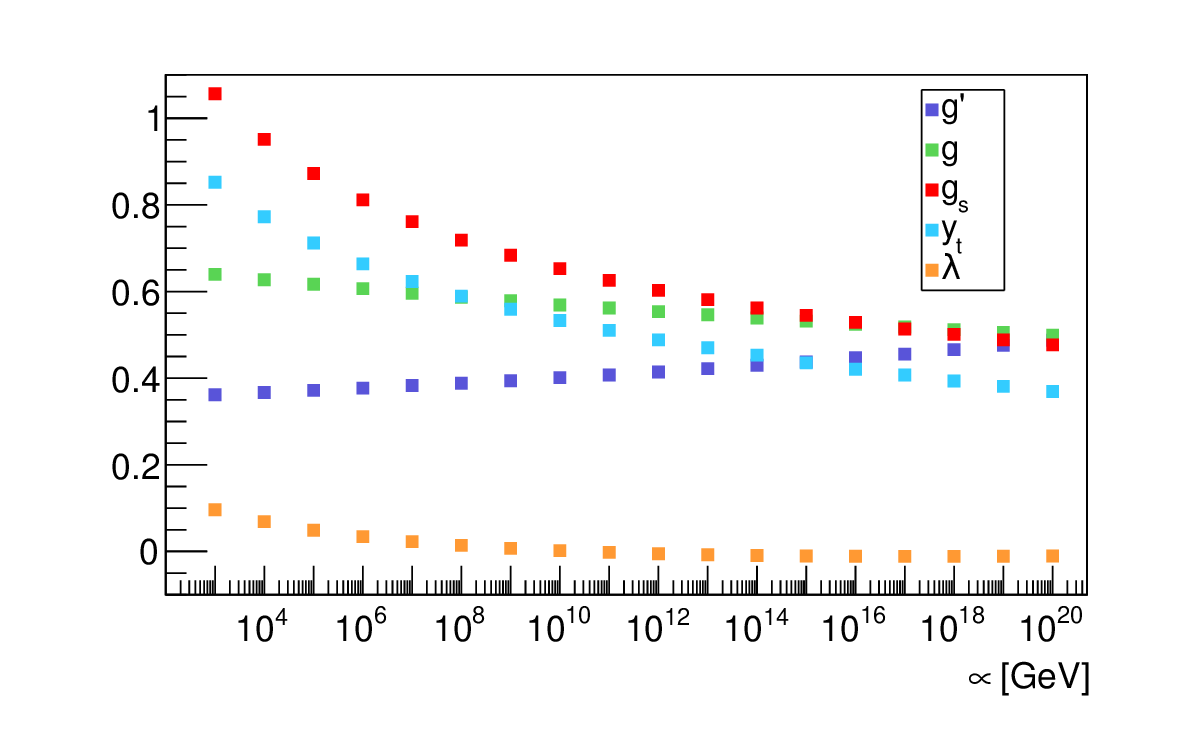}
\includegraphics[width=0.53\textwidth]
{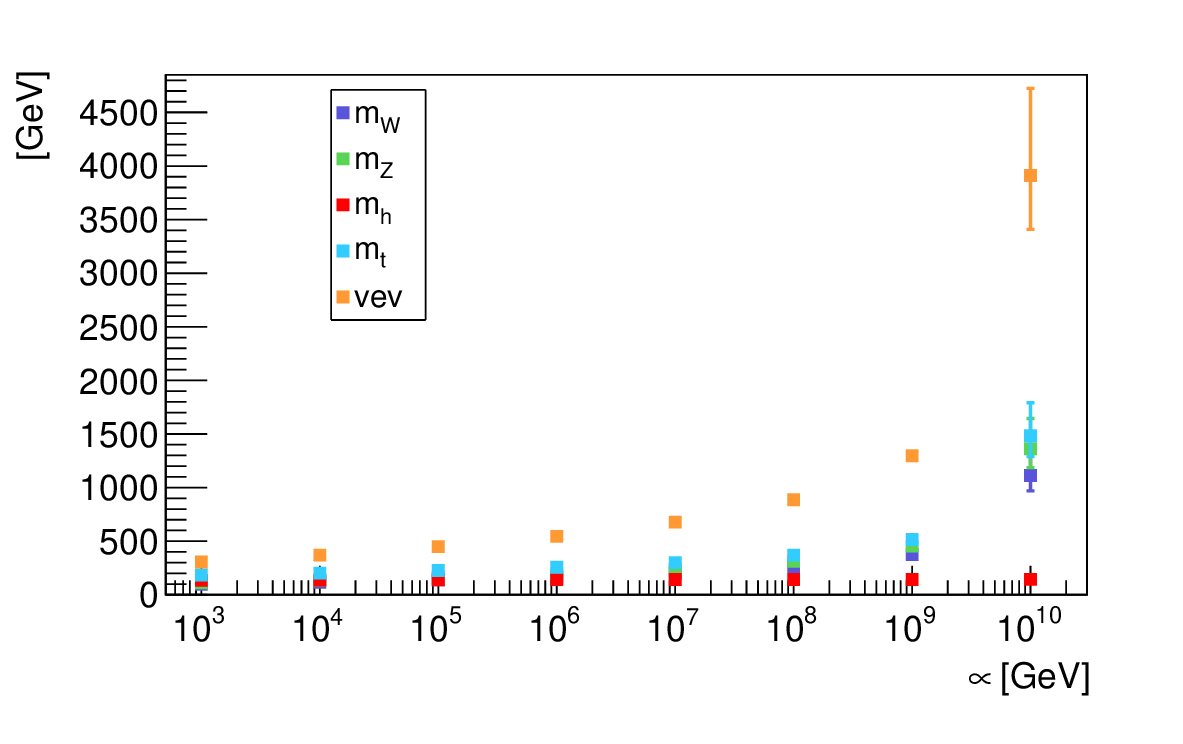}}
\caption{Left: Running of the Standard Model
gauge couplings $g$, $g'$, $g_s$
for the electroweak SU(2) and U(1) and colour SU(3), 
the top quark Yukawa coupling $y_t$ 
and Higgs self-coupling $\lambda$.
(From left, the points describe the evolution of
$g_s$, $y_t$, $g$, $g'$, $\lambda$ in descending order.)
Right: 
Running $\overline{\rm MS}$ masses and 
the Higgs vev in the Standard Model. 
Uncertainties are calculated by varying all PDG values
up and down by their respective uncertainties.
(In the printed black and white version,
 the points from top describe the evolution of
 $v$, $m_t$, $m_Z$, $m_W$, $m_h$.)
Figure taken from \cite{Bass:2020nrg}.}
\label{fig:runningC}
\end{figure}

Besides ensuring gauge invariance with the massive W and Z gauge bosons, 
the BEH mechanism with 
the measured Higgs mass
also gives good 
high energy behaviour with 
perturbative unitarity \cite{LlewellynSmith:1973yud,Bell:1973ex,Cornwall:1973tb,Cornwall:1974km}
in scattering of the massive W and Z gauge bosons 
and a renormalisable Standard Model~\cite{tHooft:1971qjg,tHooft:1972tcz,Veltman:1968ki}.
Given the Standard Model  
couplings we can extrapolate the theory up to the Planck scale using perturbative 
RG evolution.
If we assume no coupling to undiscovered new particles, 
then it remains finite and well behaved with no Landau pole singularities below the Planck scale.
That is, the Standard Model is mathematically consistent up to the Planck scale.

Further, 
the Standard Model  Higgs vacuum comes out very close to the border of stable and metastable, within about 1.3 $\sigma$ of being stable~\cite{Bednyakov:2015sca}
-- see also 
\cite{Jegerlehner:2013cta,Degrassi:2012ry,Buttazzo:2013uya}.
Vacuum stability is connected with the RG running of 
the Higgs self-coupling 
$\lambda$, with a stable vacuum for positive definite $\lambda$.
If this coupling crosses zero deep in the ultraviolet it can lead to vacuum metastability.  
Whether this happens is 
very sensitive to the values of the Higgs boson and top quark masses and to details of calculations of higher order 
radiative corrections.
Calculations are performed with the 
Standard Model evolved up to the Planck scale 
with the measured 
masses and couplings 
as input 
and 
using 3 loop RG,
2 loop matching plus 
pure QCD corrections
evaluated to 4 loops.
In general, a higher top quark mass tends to reduce $\lambda$ deep in the ultraviolet whereas a larger Higgs mass tends to increase it. 
Sensitivity to QCD corrections involving 
$\alpha_s$ 
means sensitivity also 
to the numbers of colours and active flavours and the QCD scale
$\Lambda_{\rm qcd}$. 
Both electroweak and QCD physics 
thus enter the vacuum stability calculations.

The Standard Model running couplings are shown in 
Fig.~\ref{fig:runningC} 
which displays calculations 
\cite{Bass:2020nrg}
performed 
using the C{\tt ++} 
RG evolution package 
\cite{Kniehl:2016enc}.
The QCD and electroweak SU(2) 
couplings are asymptotically free, decaying 
logarithmically 
with increasing resolution,  
and
the U(1) coupling is non-asymptotically free rising in the ultraviolet. 
These couplings almost meet in the ultraviolet but don't quite. 
The top quark Yukawa coupling decreases with increasing resolution.
With the 
measured top quark mass $m_t$
and QCD coupling $\alpha_s$,  
the Standard Model 
needs a Higgs mass $m_h$ 
bigger than about 125 GeV 
to ensure vacuum stability, 
making the Higgs particle discovered at CERN especially interesting.
(The Higgs self coupling would generate a Landau pole singularity below $M_{\rm Pl}$ if the Higgs mass were about 30\% larger 
with the measured value of $m_t$, placing a perturbative upper bound on 
the possible Higgs mass~\cite{Hambye:1996wb}.)
To illustrate sensitivity to the top quark mass,  
if we set $m_h=125$ GeV 
and vary the top quark mass in these calculations, 
then the Higgs self-coupling $\lambda$
crosses zero around $10^{10}$ GeV with $m_t=173$ GeV  
\footnote{Vacuum metastability would come with a 
new minimum in the effective Higgs potential not far below the Planck scale and a vacuum 
lifetime 
$\sim 10^{600}$ years, 
very much bigger 
than the present age of the Universe.
}
and remains positive definite
up to $M_{\rm Pl}$ 
with $m_t=171$ GeV.
The recent and most accurate  measurement of $m_t$ from CMS Run-2 data 
collected at $\sqrt{s} = 13$ TeV is 
$171.77 \pm 0.37$ GeV~\cite{CMS:2023ebf}. 
Theoretical issues connecting the top quark Yukawa coupling and the measured mass are discussed in \cite{Hoang:2020iah}.

In parallel to the running couplings, 
for illustration the right panel of Fig.~\ref{fig:runningC}
shows the RG behaviour of the Standard Model  
${\overline {\rm MS}}$ 
masses and the Higgs 
vacuum expectation value $v$.  
These are related to the running couplings through Eqs.~(\ref{eq:2a})--(\ref{eq:2c}). 
The Higgs mass $m_h$ evolves smoothly with increasing scale $\mu$.  
In Fig.~\ref{fig:runningC} the divergence of $v$ at 
$\mu \approx 10^{10}$ GeV corresponds to the  scenario where $\lambda$ crosses zero at this scale. 
With a stable vacuum 
$\lambda$ remains positive 
and $v$ is finite without divergence.

Summarising, if we assume that there are 
no couplings to extra particles at higher energies,
then the Standard Model revealed by current experiments is 
strongly correlated with its behaviour in the extreme ultraviolet.
This may be telling us 
something deeper about
the origin of the Standard Model.
It is important to emphasise the large extrapolations
in these calculations when evolving the Standard Model couplings up
to the Planck scale. 
The existence of new physics, even at the largest scales, can affect the vacuum stability \cite{Branchina:2013jra}.
Modulo this caveat, 
within the uncertainties on the top mass and radiative corrections, vacuum stability 
is certainly not excluded up to the energies where Grand Unified Theories are 
sometimes conjectured to apply 
or even up to the Planck scale 
\cite{Jegerlehner:2013cta,Bednyakov:2015sca}.
The 
parameters of the Standard Model  
might thus be 
linked to physics 
at these high scales.  
In particular, the Higgs mass 
might be environmentally determined, linked to the stability of the electroweak vacuum.
The Higgs vacuum sitting "close to the edge" 
of stable and metastable may be pointing to 
possible new critical phenomena in the ultraviolet~\cite{Jegerlehner:2013cta,Degrassi:2012ry,Buttazzo:2013uya}. 
There are ideas that 
near-criticality 
might act as an attractor point in 
evolution of the higher energy phase, 
with analogous systems in Nature discussed in ~\cite{Buttazzo:2013uya}.
With just small changes in the Standard Model parameters the emerging low energy theory would be very different from the Standard Model.

If the Standard Model might really work up to these 
large scales, 
then how might we understand the origin of gauge symmetries and 
the new physics needed to the explain the open puzzles at the interface of particle physics and cosmology?

\section{Emergent gauge symmetries}
\label{sec:EGS}

Gauge symmetries 
can be "emergent" below a scale deep in the ultraviolet.

Emergence in physics occurs when a many-body system exhibits collective behaviour in the infrared
that is qualitatively different from that of its more primordial constituents as probed in the ultraviolet \cite{Anderson:1972pca,Palacios:book}.
For example, hadrons are emergent from quark gluon degrees of freedom. 
As an
everyday example of emergent symmetry, consider a carpet which looks flat and translational
invariant when looked at from a distance. Up close, e.g. as perceived by an ant crawling on it,
the carpet has structure and this translational invariance is lost. The symmetry perceived in the
infrared, e.g. by someone looking at it from a distance, “dissolves” in the ultraviolet when the
carpet is observed close up.

Emergent gauge symmetries are seen in 
quantum many-body systems 
where they are associated with 
topological phase transitions, e.g., 
phase transitions 
that occur without an order parameter 
\cite{Zaanen:2011hm,Sachdev:2015slk,Levin:2004js,Volovik:2008dd,
Volovik:2003fe,Affleck:1988zz,Baskaran:1987my}.
Here, 
the prototype is the Fermi-Hubbard model of 
strongly correlated electrons
in an atomic lattice 
which exhibits spin-charge separation with 
the SU(2) spin becoming dynamical 
with an emergent gauge symmetry beyond the more fundamental QED~\cite{Baskaran:1987my,Affleck:1988zz}. 
A further example is 
the A-phase of superfluid $^3$He 
\cite{Volovik:2003fe,Volovik:2008dd}.
Close to a Fermi point one finds emergent local 
gauge interactions with 
spin becoming  dynamical 
to internal observers. 
The quasi-particles include gauge bosons and also 
emergent chiral fermions, 
each with common 
limiting velocity 
like with Lorentz invariance in the Standard Model, 
and an emergent metric as well as an 
analogue of the 
chiral anomaly.
Emergent gauge symmetries 
are also 
believed to be important in 
high temperature superconductors 
\cite{Sachdev:2015slk}.

How might emergent gauge symmetries work in particle physics and might the Standard Model be emergent?

Consider a statistical system near its critical point.
The long range asymptote is a renormalisable quantum field theory with properties described by the renormalisation group
\cite{Wilson:1973jj,Peskin:1995ev}.
If the spectrum includes 
J=1 excitations among 
the degrees of freedom in the low energy phase, 
then it is a gauge theory. 
Gauge symmetries would then be an emergent property of the low energy phase and "dissolve" in some topological like
phase transition
deep in the ultraviolet 
~\cite{Jegerlehner:1978nk,Jegerlehner:1998kt,Jegerlehner:2013cta,Forster:1980dg,tHooft:2007nis,Bjorken:2001pe,Bass:2021wxv}.
The quarks and leptons as well as the gauge bosons and
Higgs boson would then be the stable collective long-range excitations of some (unknown) 
more primordial 
degrees of freedom that exist above the scale of emergence.
The vacuum of the low energy phase should be stable below the scale of emergence.

Suppose the Standard Model works like this.
Then it behaves like an 
effective theory with the theory at mass dimension four, $D=4$,  supplemented by a tower of non-renormalisable  higher dimensional operators suppressed by factors of the large scale of emergence. 
The global symmetries of the Standard Model at $D=4$ 
are constrained by gauge invariance and renormalisability.
The higher dimensional operators are less constrained 
and may exhibit extra global symmetry breaking \cite{Weinberg:2018apv,Jegerlehner:2013cta,Bass:2021wxv,Bass:2020gpp,Witten:2017hdv}.
Lepton number violation and 
tiny Majorana neutrino masses may enter at $D=5$, 
suppressed by single power of
the scale of emergence, 
via the so called Weinberg operator  \cite{Weinberg:1979sa}. 
One finds 
$m_\nu \sim \Lambda_{\rm ew}^2/M$ 
with $\Lambda_{\rm ew}$ 
the electroweak scale 
and 
$M$ the scale of emergence. 
Possible proton decays might occur at $D=6$ suppressed by two powers of $M$ 
\cite{Weinberg:1979sa,Wilczek:1979hc}.
New CP violation, needed for baryogenesis, 
might occur 
in Majorana phases at $D=5$ 
as well as in 
new $D=6+$ operators 
\cite{Grzadkowski:2010es}.
\footnote{LHC data (so far) reveal no evidence for higher dimensional correlations divided by powers of a large mass scale below the few TeV range \cite{Slade:2019bjo,Ellis:2020unq}.}

If one increases the energy much above the
electroweak scale, then the physics becomes increasingly  symmetric with energies 
$E \gg \Lambda_{\rm ew}$ 
until we come  
within about 0.1\% 
or so of the scale of emergence. Then new global symmetry violations in higher dimensional operators become important so the physics becomes increasingly chaotic until one goes through the 
phase transition associated with the scale of emergence,
with the physics above this scale 
then described 
by new degrees of freedom and perhaps new physical laws.
This scenario contrasts with unification models which involve maximum symmetry 
in the extreme ultraviolet. 
The effect of higher dimensional operators might have been especially active in the very early Universe with energies $E \sim M$ and be manifest through tiny numbers today.

In other ideas,  
emergent gauge symmetries can also appear through decoupling of gauge violating terms 
in the RG at an infrared fixed point \cite{Wetterich:2016qee}
and in connection with possible spontaneous breaking of Lorentz symmetry, SBLS 
\cite{Bjorken:2001pe,Bjorken:2010qx,Bjorken:1963vg,Chkareuli:2001xe}.
In the former case, 
the coefficient of any 
local gauge symmetry violating terms 
blows up at the fixed point, 
in contrast to restoration of global symmetries where the coefficient of any symmetry violating term goes to zero at the fixed point. 
With SBLS, 
non observability of 
any Lorentz violating terms at $D=4$ 
corresponds to 
gauge symmetries 
for vector fields like the photon. 
Possible Lorentz violation might be manifest at 
${\cal O} (\Lambda_{\rm ew}^2/M^2)$ 
with a 
preferred reference frame naturally identified with 
the frame where the cosmic microwave background
is locally at rest 
\cite{Bjorken:2001pe}.

When coupling to gravitation, this emergence picture 
gives a simple 
explanation of the 
cosmological constant.
One also finds interesting constraints on possible dark matter scenarios.

\subsection{Vacuum energy and the cosmological constant}

The simplest explanation of the accelerating expansion of the Universe is a 
cosmological constant $\Lambda$ 
in Einstein's equations of General Relativity, 
\begin{equation}
R_{\mu \nu} - \frac{1}{2} g_{\mu \nu} \ R = 
- \frac{8 \pi G}{c^4} T_{\mu \nu} + \Lambda g_{\mu \nu} .
\label{eq3.1}
\end{equation}
Here $R_{\mu \nu}$ is the Ricci tensor, 
$R$ is the Ricci scalar
and 
$T_{\mu \nu}$ is the energy-momentum tensor 
for excitations above the vacuum;
$G$ is Newton's constant and $c$ is the speed of light. 
\footnote{Here we follow 
the convention 
with the Minkowski metric taken as  diag[-1,+1,+1,+1] 
so that positive $\Lambda$
corresponds to a 
positive energy density and a negative pressure
\cite{Weinberg:1988cp}.}
Matter 
including dark matter 
clumps together under normal gravitational
attraction whereas the cosmological constant 
is the same at all points in space-time. 
Einstein's equations determine the geodesics on which particles
propagate in curved space-time in the presence of a gravitational source.

The cosmological constant 
measures the vacuum energy density perceived by gravitation,
\begin{equation}
\rho_{\rm vac} = \Lambda 
\times 
 c^4 
/ (8 \pi G ) ,
\label{eq3.2}
\end{equation}
with 
associated scale $\mu_{\rm vac}$,
$\rho_{\rm vac} = \mu_{\rm vac}^4$.
Astrophysics observations \cite{Planck:2018vyg} 
tell us 
that $\Lambda = 1.088 \times 10^{-56} \ {\rm cm}^{-2}$ corresponding to 
\begin{equation}
\rho_{\rm vac} = (0.002 {\rm \ eV})^4
\label{eq3.3}
\end{equation}
with a present period of accelerating expansion 
that began about five billion years ago when the matter density of the expanding Universe fell below 
$\rho_{\rm vac}$, which then 
took over as the main  driving term.
So far, astrophysics data is consistent with 
dark energy being 
a time independent cosmological constant
with vacuum equation of state 
(energy density = - pressure)~\cite{Escamilla:2023oce}.

The small value 0.002 eV 
in Eq.~(\ref{eq3.3}) is intriguing from the viewpoint of Standard Model 
quantum fields.
In general, vacuum energy is sensitive to quantum fluctuations and potentials in the vacuum with terms involving the much larger 
QCD and electroweak  scales ($\approx 200$ MeV and 246 GeV). 
If taken alone, 
these particle physics contributions 
give large curvature contributions 
when substituted into Einstein's equations -- inconsistent with the 
flat Universe we observe in cosmology. 
This issue for zero-point energies was noticed already in the early works \cite{Pauli:1933b,Zeldovich:1967gd} 
with spontaneous symmetry and vacuum potentials discussed in  \cite{Dreitlein:1974sa,Veltman:1997nm}. 
(The history of thinking on this topic is reviewed in  \cite{Straumann:2002tv,Kragh:2014jaa}.)
The net 
$\rho_{\rm vac}$ 
corresponding to the cosmological constant 
also involves   
an extra gravitational contribution 
which may be dynamical 
-- for detailed discussion see Section \ref{sec:CC}.
Vacuum energy only becomes an observable when coupling to gravitation via the cosmological constant. 
Without gravitational coupling
only energy differences count, 
so in usual particle physics one is free to set the energy of the vacuum to zero 
- e.g., through normal ordering (before considerations of spontaneous symmetry breaking). 
The cosmological constant 
is an observable measurable through the accelerating expansion of the Universe.
Hence it should 
be particle physics 
RG scale invariant.
Individual contributions are, in general, 
RG scale dependent 
(so theorist/calculation dependent)
with only the net 
cosmological constant 
as the observable. 
Why is the cosmological constant so small?

The cosmological constant is connected with the symmetries of the metric. 
With a finite cosmological constant Einstein's equations
have no vacuum solution 
where $g_{\mu \nu}$ is the constant Minkowski metric.
That is, 
global space-time translational invariance 
(a subgroup of the group of general co-ordinate transformations)
of the vacuum is broken 
by a finite value of $\rho_{\rm vac}$
\cite{Weinberg:1988cp}.
\footnote{
Set $T_{\mu \nu}=0$ to remove excitations above the vacuum. 
If the global Minkowski metric is a solution,
the on the left hand side 
of Eq.~(\ref{eq3.1}) 
there is no curvature, 
derivatives will vanish 
with $R=0$ 
so the cosmological constant will also vanish.}
The reason is that a finite value of 
$\rho_{\rm vac}$ acts as a gravitational source   
 which generates a dynamical space-time  
 with accelerating expansion for positive $\rho_{\rm vac}$.
Suppose the vacuum including condensates with finite vacuum expectation values 
is space-time translational invariant 
and that  
flat space-time is consistent at mass  dimension four, 
just as suggested by the success of the Standard Model.
With the Standard Model as an effective theory emerging 
in the infrared,
the low-energy global symmetries 
including space-time translation invariance
can be broken through additional higher dimensional terms,
suppressed by powers of the
large scale of emergence $M$.
Then 
the RG invariant scales
$\Lambda_{\rm qcd}$ and 
electroweak $\Lambda_{\rm ew}$
might enter the cosmological constant
with the scale of the leading term suppressed by 
$\Lambda_{\rm ew}/M$ 
(that is, $\rho_{\rm vac} \sim (\Lambda_{\rm ew}^2/M)^4$
 with one factor of 
 $\Lambda_{\rm ew}^2/M$ for each dimension of space-time)
 -- see Refs.~\cite{Bass:2020egf,Bass:2020nrg} and the early work
\cite{Bjorken:2001pe,Bjorken:2001yv}.
This scenario, if manifest in nature, 
would explain why the cosmological constant scale 0.002 eV 
is similar to what we expect for the neutrino masses 
\cite{Altarelli:2004cp}
\footnote{
Assuming three species of neutrinos,
the neutrino oscillation data 
constrains the largest 
mass squared difference to be
$\approx 2 \times 10^{-3}$ eV$^2$ 
with the smaller one as 
$(7.53 \pm 0.18) \times 10^{-5}$ eV$^2$
\cite{BahaBalantekin:2018ppj}.
}, 
which for Majorana neutrinos are
themselves linked to the dimension five Weinberg 
operator with
$m_{\nu} \sim \Lambda_{\rm ew}^2/M$ \cite{Weinberg:1979sa}.
That is, 
\begin{equation}
 \mu_{\rm vac} \sim 
 m_\nu \sim \Lambda_{\rm ew}^2/M   .
\label{eq3.4}
\end{equation}
Here one is taking the Standard Model as describing 
particle interactions at $D=4$ up to the large  scale $M$.
The cosmological constant would vanish at  mass dimension four.
This vanishing cosmological constant contribution 
is equivalent
to a renormalisation condition $\rho_{\rm vac} =0$
at $D=4$ 
imposed by  global space-time translational
invariance of the vacuum, 
even in the presence of  QCD and Higgs condensates.
The precision of global symmetries in our experiments,
e.g., 
lepton and baryon number conservation, 
tells us that in this scenario the scale of emergence should 
be deep in the ultraviolet, 
much above the Higgs and other Standard Model particle masses. 
Taking the value 
$\mu_{\rm vac} = 0.002 {\rm \ eV}$
from astrophysics 
together with $\Lambda_{\rm ew} = 246$ GeV 
gives a value for $M$ about $10^{16}$ GeV.

The scale $10^{16}$ GeV is within the range where 
the Higgs self-coupling 
$\lambda$ might cross zero (if indeed it does) under RG evolution.
It is also similar to the "GUT scale" 
that typically  appears in unification models.
If the Standard Model is emergent at a scale where $\lambda$ 
is non-negative then its vacuum will be fully  stable.
Any perturbative extrapolation of Standard Model degrees of freedom above the scale of emergence would reach into an unphysical region.

While there is no evidence in present data for a time dependent dark energy, 
the search for possible time dependence as well as any deviations from General Relativity is proceeding with  next generation cosmological surveys, e.g. with EUCLID, 
the JWST, ELT, and the proposed Einstein telescope. 
As an example, 
de Sitter space is unstable in the 
$S-$matrix formulation of quantum gravity 
with a quantum breaking effect and (slow) decay of the cosmological constant~\cite{Dvali:2020etd,Dvali:2017eba}.
Within the emergence scenario 
any time dependence 
might 
correspond either to a change in 
$\Lambda_{\rm ew}^2/M$ 
(with $G$ and $c$ taken as fixed)
or to a vanishing cosmological constant beyond leading order in $1/M$ and dark energy instead 
corresponding to some new dynamics with just gravitational coupling, 
e.g., some BRST invariant coherent state of gravitons as suggested in  \cite{Berezhiani:2021zst}.
In the latter case, 
one then has the phenomenological issue of explaining the 
similar sizes   
$\mu_{\rm vac} \sim m_\nu$ observed today. 
\footnote{Other ideas relating the 
sizes of the cosmological constant scale and neutrino mass are discussed in  
Refs.~\cite{Wetterich:2007kr,Brookfield:2005bz,Fardon:2003eh}.}

Cosmology observations of a spatially flat Universe
\cite{Planck:2018vyg} 
constrain the energy densities of dark energy and dark matter.
For a flat Universe 
the density parameter
$\Omega = 8 \pi G \rho / 3H_0^2 =1$ 
where $\rho$ 
is the sum over individual 
matter, radiation and dark energy density contributions.
The normal baryonic 
and cold dark matter, radiation and dark energy contributions 
 measured by Planck are  
 $\Omega_b = 0.049$,
 $\Omega_{\rm cdm} = 0.265(7)$, 
 $\Omega_\gamma = 2.5 \times 10^{-5}$, 
 $\Omega_\Lambda = 0.685(7)$ 
 with the 
 Hubble constant today 
 $H_0 = 0.674(5) \
 {\rm km s^{-1} Mpc^{-1}}$ 
 \cite{Planck:2018vyg}
 \footnote{This measurement of $H_0$
 from the cosmic microwave background  
 is in tension with 
 recent late-time, 
 low-redshift measurements \cite{Riess:2019qba}  
 which give a value of 
 $H_0$ about $0.73 \ {\rm km s^{-1} Mpc^{-1}}$
 -- a present open puzzle in cosmology.}.
The curvature density parameter comes out as 
$\Omega_k = 1 - \Omega = k/{\dot a}^2 = 0.0007 (19)$ 
with $k$ the curvature and $a$ the scale factor appearing in the 
Friedmann–Lemaitre–Robertson–Walker, FLRW, metric 
($H_0 = {\dot a}/a$ with $a=1$ taken today) 
suggesting a flat Universe.  
Observation of a spatially flat Universe together with measurements of the 
dark matter contribution 
suggested the presence of a finite cosmological constant 
\cite{Efstathiou:1990xe}
before the discovery of accelerating expansion of the Universe through supernova Sn1a observations 
\cite{SupernovaSearchTeam:1998fmf,SupernovaCosmologyProject:1998vns}.
Alternatively, with the 
cosmological constant scale deduced using the emergent Standard Model arguments above, 
Eq.~(\ref{eq3.4}), 
flatness 
gives a constraint on the amount of extra dark matter  needing extra theoretical understanding. 

The observations of 
flatness, 
isotropy and homogeneity plus today the lack of magnetic monopoles which might have been produced in the very early Universe 
motivate primordial inflation ideas 
\cite{Baumann:2008bn,Ellis:PDG}
where initial exponential expansion is driven by an 
inflaton scalar field and  slow-roll potential condition. 
In this picture 
the scale of inflation is typically taken as around 
$10^{16}$ GeV, 
which is 
close to $M$, our scale of emergence.
Perhaps the physics underlying inflation and 
emergence might be connected.
The size of the 
cosmological constant
is further constrained by anthropic arguments, 
that if
$\Lambda$ were more than about ten times larger, galaxies would not have had time to form~\cite{Weinberg:1987dv}.

\subsection{Dark matter and higher dimensional 
couplings}

Some extra 
stable dark matter
is suggested by studies of galaxies and galaxy clusters, 
gravitational lensing and the cosmic microwave background 
\cite{Baudis:2018bvr,Bertone:2018krk,Wechsler:2018pic}.
We see just a 5\%
contribution to the energy budget of the Universe 
in visible matter; 
68\% is in dark energy and 27\% in dark matter observed (so far) just through gravitational effects.
The quest to understand 
this dark matter has inspired vast experimental and theoretical 
activity with ideas including possible 
new types of elementary particles, primordial black holes as well as
possible extensions of General Relativity.

The mass scale of 
possible new dark matter particles 
is so far 
not known from experiments with ideas including new particles with masses ranging from 
$10^{-22}$ eV up to 
$10^{15}$ GeV 
-- a range of 10$^{46}$ 
\cite{Baudis:2018bvr}.
If the Standard Model  
does describe 
particle physics
interactions at $D=4$ 
up to the highest scales,
then we do not expect lightest mass 
supersymmetry 
particles or  
extra inert Higgs doublet states as sources of dark matter. 
With emergence small multiplets are preferred 
as collective excitations of the system that resides deep in the ultraviolet \cite{Jegerlehner:2013cta}.
One might then look to terms suppressed by powers of $1/M$, whether new particle couplings or 
modified gravity scenarios, 
e.g. \cite{Verlinde:2016toy}.
As a particle candidate, 
if present, pseudoscalar axions~\cite{Kawasaki:2013ae}  would enter at $D=5$ with masses and couplings to Standard Model particles suppressed by a single power of a large axion "decay constant",  
which might be associated with any scale of emergence.~\footnote{Axions are a candidate
to explain the strong CP puzzle (why gluon topological effects do not induce extra CP violation in QCD) \cite{Peccei:1977hh,Weinberg:1977ma,Wilczek:1977pj}
though this might also be associated with 
confinement related phenomena~\cite{Nakamura:2021meh}.  
QCD axion masses are usually taken 
from astrophysics and cosmology constraints 
between 
1 $\mu$eV and 3 meV corresponding to an ultraviolet scale between 
$6 \times 10^9$ and 
$6 \times 10^{12}$ GeV.  This is less than the $10^{16]}$ GeV 
discussed here in connection with the cosmological constant and neutrino masses. 
Possible relaxation of the axion bounds to include perhaps higher scales is discussed in \cite{Dvali:1995ce}.
Any 
axion 
Bose-Einstein condensate 
contribution would 
approximately scale  
as $D=4$ 
with the axion decay constant 
suppression 
factor missing in the 
net condensate contribution \cite{Sikivie:2009qn}.
}

Besides possible new particles, 
primordial black holes are discussed as candidates 
for dark matter~\cite{Carr:2021bzv,Green:2020jor},  
and are consistent with General Relativity with coupling up to $D=4$.
Suggestions include the black holes in the mass range 
observed with the
LIGO-VIRGO 
gravitational wave measurements~\cite{Boehm:2020jwd}.
Primordial black holes 
formed in the early Universe 
before the time of 
stellar collapse origin 
might be detected 
in future gravity waves measurements  
\cite{Bertone:2019irm}.

Whilst we search for evidence of possible new dark matter particles and/or 
black holes formed in the early Universe 
it is worthwhile also to consider 
other observables that might give clues to the nature of dark matter.
Interesting information may follow from studying the redshift dependence of 
correlations between 
dark matter and galaxy properties, 
and also 
common 
trajectories followed by particle signals from distant galaxies. 
One observes phenomenological
correlations between 
galaxy rotation velocities 
and dark matter 
\cite{Wechsler:2018pic}
and correlations 
between the masses of 
supermassive black holes  at galaxy centres and their dark matter halos 
\cite{Ferrarese:2002ct,Ferrarese:2004qr}.
These correlations have 
so far been measured  
just for galaxies at small redshifts.
New windows of observation are now opening up with instruments such as,  
e.g., the JWST, ELT...

The baryonic version of of the Tully-Fisher relation connects the total baryonic mass in a galaxy 
${\cal M}_{\rm bar}$
to the asymptotic rotation velocity $v_{\infty}$
\cite{McGaugh:2000sr},
\begin{equation}
v_{\infty}^4 = G {\cal M}_{\rm bar} a_0 , 
\label{eq3.5}
\end{equation}
where $a_0$ is a phenomenological 
acceleration parameter,
$
a_0 \approx 1.2 \times 10^{-10} \ {\rm m s^{-2}}
$
to within 10-20\%  \cite{Milgrom:2020cch}
obtained from galaxy rotation curves.
Numerically 
$2 \pi a_0 
\approx a_H 
\approx a_{\Lambda}
$
where
$a_H = {c H_0} \approx 1.1 \times 10^{-10}$ ms$^{-2}$
and
$a_{\Lambda} = c^2 \sqrt{\Lambda /3}
=
c H_{\infty} \approx 0.9 
\times 10^{-10}$ ms$^{-2}$.
Here $H_0$ is the Hubble parameter today, 
$\Lambda$
is the cosmological constant and
$H_\infty$ is the Hubble constant in the infinite
future of $\Lambda$CDM
after matter and radiation
density terms have decayed 
approaching zero with
the cosmological constant taken as time independent.
Eq.(\ref{eq3.5}) is sometimes taken as phenomenological evidence to hint that usual Newtonian gravity might be modified, the so called MOND theories \cite{Milgrom:1983pn}.
\footnote{
These involve an extra asymptotic 
long-range gravitational acceleration 
$g = a_0 \frac{r_M}{r}$ with $r_M = ({\cal M}G/a_0)^{\frac{1}{2}}$ 
where 
${\cal M}$ is the central mass.
}
It may also be
the net result that should
come out from simulating a 
microscopic theory of dark matter involving 
new dark matter particles
within 
usual General Relativity and connected to galaxy formation.
This issue is a step beyond the simple fit to data associated with
Eq.(\ref{eq3.5}).
Here we just note that
$a_H$ and $a_{\Lambda}$
have different scaling behaviour in a $1/M$ expansion with $\mu_{\rm vac} \sim \Lambda_{\rm ew}^2/M$: 
$a_H$ has a mass dimension four component whereas $a_{\Lambda} \sim 1/M^2$.
Further, $a_H$ is time dependent through $H_0$
whereas $a_{\Lambda}$ is time independent if we take a time independent cosmological constant.
It would be interesting to measure the redshift dependence of the phenomenological relation in  Eq.(\ref{eq3.5}) 
through 
more precise
observations of distant galaxies: might the value of $a_0$ be time independent or its similarity with the Hubble constant 
just a coincidence at low redshift?
Possible redshift dependence of $a_0$, 
including with $\Lambda$CDM dynamics, 
is discussed in 
\cite{Mayer:2022qhk}.
If $a_0$ were found to be redshift independent,
then it might hint at dark matter being associated 
with some higher dimensional coupling involving 
the cosmological constant.

Another interesting 
constraint is that the gravitational effect of dark matter or any modified gravity scenarios 
is the same for all particles. 
Whatever dark matter is, 
within present experimental 
uncertainties it does not seem to affect the common trajectories  travelled by gravitational waves and by different particle species~\cite{Baudis:2018bvr}.
In multimessenger astronomy 
light and gravitational wave signals 
identified with 
the neutron star merger event 
GKB170817 were 
observed to arrive very close to  simultaneously   
constraining the speeds of gravity and light to be the same to within $\sim 10^{-15}$ times 
the speed of light, 
meaning that they follow the same  geodesics~\cite{LIGOScientific:2017zic}. 
Likewise, the initial neutrino multimessenger event 
IceCube-170922A
with energy 290 TeV \cite{IceCube:2018dnn} 
and the neutrino event 
Icecube-141209A
with $E_\nu=97.4 \pm 9.6$ TeV
detected in coincidence with the  blazar 
GB6 J1040+0617
in a phase of high $\gamma$-ray activity \cite{Kun:2023uld}
each correspond to 
the photon and neutrino speeds 
being the same to within about $10^{-11}$ of the speed of light 
assuming the neutrinos were emitted coincident with the blazar flaring periods.
\footnote{
The gravitational wave 
event GKB170817 was observed  
in coincidence  with a 
short $\gamma$-ray burst,
GRB 170817A at redshift
$z= 0.008^{+0.002}_{-0.003}$, 
with signals detected 2 seconds apart originating
from a neutron star merger event
more than $10^8$ light years away.
The neutrino event IceCube-170922A 
was identified in parallel 
with very high energy $\gamma$-rays from the blazar source TXS 0506+056 at redshift $z=0.3365 \pm 0.0010$ 
with the blazar 
active within about $\pm 14$ days of the 290 TeV 
neutrino event.
The statistically favoured joint blazar excitation 
and neutrino event
IceCube-141209A 
was observed in a window of high 
$\gamma$-ray activity from the blazar GB6 J1040+0617 at redshift 
$z=0.7351 \pm 0.0045$ 
which was brightest 4.5 days before the neutrino detection.
}

Different particle trajectories might be induced in exotic 
modified gravity scenarios with different metrics 
for gravitational waves 
coupling to General Relativity without dark matter
and 
for normal matter coupling to General Relativity with dark matter. 
Trajectories might also be affected by any 
new direct matter to curvature couplings which, if present, 
would involve an extended
gravitational action involving a series in 
${\cal O}(R/M^2)$, 
e.g. involving terms like 
$1/M^2 R_{\mu \nu} F^{\mu \omega} F^{\nu}_{ \ \omega}$ with 
$F_{\mu \nu}$ 
the electromagnetic field tensor.
Such terms preserve the 
local gravitational symmetries of invariance
under local co-ordinate transformations 
but change the dynamics  \cite{Charlton:2020kie,Shore:2004sh}.
They modify the 
energy momentum tensor and equations of motion and 
can lead to particles not following geodesics,
with the effect in general
different for different particles. 
If present, these terms would contribute in regions of strong gravitational curvature. 
In addition, 
possible violations of Lorentz invariance discussed in 
\cite{Bjorken:2001pe}
might arise at 
${\cal O}(\mu_{\rm vac}/M \sim \Lambda_{\rm ew}^2/M^2)  \approx 10^{-28}$.

\section{Structure of the cosmological constant}
\label{sec:CC}

So far we have addressed the cosmological constant within the framework of an emergent Standard Model.
The tiny cosmological constant scale 0.002 eV 
is very much smaller than the electroweak scale 246 GeV which is, in turn, very much smaller than the Planck scale.
In this Section and Section \ref{sec:Scales} 
we discuss these scale hierarchies and how they connect with issues of renormalisation and ultraviolet regularisation.

The cosmological constant through $\rho_{\rm vac}$ 
receives contributions 
from the zero point energies, ZPEs, 
of quantum field theory
through vacuum-to-vacuum diagrams  \cite{Jaffe:2005vp},
any (dynamically generated) potential in the vacuum, 
e.g. induced by spontaneous symmetry breaking and Higgs and QCD condensates~\cite{Dreitlein:1974sa,Veltman:1997nm},
and the renormalised version of a
bare gravitational term 
$\rho_{\Lambda}$, 
viz.~\cite{Weinberg:1988cp,Sola:2013gha} 
\begin{equation}
\rho_{\rm vac} = \rho_{\rm zpe} + \rho_{\rm potential} 
+ \rho_{\rm \Lambda}.
\label{eq4.1}
\end{equation}
As an observable 
the cosmological constant
is renormalisation scale invariant.
It is independent of how a theoretician might choose to calculate it,  
\begin{equation}
\frac{d}{d \mu^2} \rho_{\rm vac} =0 . 
\label{eq4.2}
\end{equation}
(Here we take Newton's constant  
$G$ as RG scale invariant with gravity treated as classical.) 
Individual terms in 
Eq.(\ref{eq4.1}) contributing to 
$\rho_{\rm vac}$ are, however,
renormalisation scale dependent as well as sensitive to large particle physics scales. 
The Higgs potential is RG scale dependent through the Higgs 
self-coupling,   
which determines the stability of the electroweak vacuum.
The ZPEs are discussed in detail below.
This renormalisation scale dependence should cancel 
with the gravitational 
term $\rho_\Lambda$
to give the scale invariant $\rho_{\rm vac}$.
The important question then is what is left over   
or 
how the different terms combine.
The small cosmological constant and net 
$\rho_{\rm vac}$  
are determined by the symmetries of spacetime 
with Nature 
so much liking the Minkowski metric and a spatially flat Universe today.
In any self-consistent discussion, each of the three terms in Eq.~(\ref{eq4.1}) should be defined with respect to the same renormalisation scheme.
The net $\rho_{\rm vac}$ obeys the vacuum equation of state, EoS.

ZPEs are a vacuum energy contribution 
induced by quantisation 
and are an intrinsic part of quantum field theory.  
One sums over harmonic oscillator contributions.
When we evaluate the ZPEs, 
they come with an ultraviolet divergence requiring regularisation and renormalisation.

Working in flat space-time the ZPE for a particle with mass $m$ is 
\begin{equation}
\rho_{\rm zpe} 
=
\frac{1}{2}
\sum \{ \hbar \omega \}
=
\frac{1}{2} \hbar
\sum_{\rm particles} g_i \int_0^{k_{\rm max}}
\frac{d^3 k}{(2 \pi)^3} \sqrt{k^2 + m^2}
.
\label{eq:4c}
\end{equation}
Here 
$m$ is the particle mass;
$g_i = (-1)^{2j} (2j+1) f$
is the degeneracy factor for a particle $i$ of spin $j$, 
with $g_i  >0$ for bosons and $g_i < 0$ for fermions.
The minus sign follows from the Pauli exclusion principle and 
the anti-commutator relations for fermions.
The factor $f$ is 1 for bosons, 
2 for each charged lepton
and 6 for each flavour of quark
(2 charge factors for the quark and antiquark, 
 each with 3 colours).
The corresponding vacuum pressure is \cite{Martin:2012bt} 
\begin{equation}
p_{\rm zpe} =
\frac{1}{3} 
\frac{1}{2} \hbar
\sum_{\rm particles} g_i \int_0^{k_{\rm max}}
\frac{d^3 k}{(2 \pi)^3} \frac{k^2}{\sqrt{k^2 + m^2}} .
\label{eq:4d}
\end{equation}
When evaluating the integrals in 
Eqs.~(\ref{eq:4c}) and (\ref{eq:4d}) 
one finds that
the equation of state is sensitive to the choice of ultraviolet regularisation.
A Lorentz covariant 
regularisation is needed  
to ensure that the ZPEs satisfy  
the vacuum EoS $\rho = -p$.

For example, suppose we evaluate the ZPE using a 
(non covariant) 
brute force 
cut-off on the 
divergent integral
with some fixed $k_{\rm max}$.  
Then the leading term in the ZPE,
which is proportional 
to $k_{\rm max}^4$, 
obeys instead 
the radiation equation of state $\rho = 3p$, viz. 
\begin{equation}
\rho_{\rm zpe}\bigg|_{k_{\rm max}} = \hbar \ g_i \ \frac{k_{\rm max}^4}{16 \pi^2}  
\biggl[ 1 + \frac{m^2}{k_{\rm max}^2} + ... \biggr], 
\ \ \ \
p_{\rm zpe}\bigg|_{k_{\rm max}} = 
\frac{1}{3} 
\hbar \ g_i \ \frac{k_{\rm max}^4}{16 \pi^2}  
\biggl[ 1 - \frac{m^2}{k_{\rm max}^2} + ... \biggr]
\label{eq:4e}
\end{equation}
with 
quadratic terms behaving as $\rho = -3 p$ and  
just subleading logarithmic terms obeying the vacuum EoS 
\cite{Martin:2012bt}.
Cosmology observations and 
local energy conservation requires that 
the net vacuum energy contribution $\rho_{\rm vac}$ obeys the vacuum EoS~\cite{Peebles:2002gy}.
While the leading term in Eq.~(\ref{eq:4e}) behaves like an homogeneous sea of radiation, it is important to recall that $\rho_{\rm zpe}$ comes from calculating vacuum-to-vacuum closed loop diagrams instead of freely propagating photons like which dominated in the early Universe.

The vacuum equation of state 
$\rho = -p$
is obtained with 
dimensional regularisation and 
minimal subtraction, 
$\overline{\rm MS}$.
Here, the ultraviolet divergence is controlled by 
analytic continuation of the theory 
including the 
dimensionality 
of loop integrals in 
the complex plane and then taking the limit that 
$D \to 4$~\cite{tHooft:1972tcz}.
One finds 
\cite{Brown:1992db,Martin:2012bt} 
\begin{equation}
\rho_{\rm zpe} = - p_{\rm zpe}
=
- 
\hbar \ g_i \
\frac{m^4}{64 \pi^2} 
\biggl[ \frac{2}{\epsilon} + \frac{3}{2} - \gamma
- \ln \biggl( \frac{m^2}{4 \pi \mu^2} \biggr) \biggr]
+ ...
\label{eq4.5}
\end{equation}
for particles with mass $m$. 
Here 
$D=4-\epsilon$ 
is the number of dimensions 
with divergence at
$D \to 4, \ \epsilon \to 0$ and
$\gamma = 0.577...$ is Euler's constant. 
The renormalisation scale $\mu$ 
enters also through the running mass and 
should decouple from the net $\rho_{\rm vac}$.
It enters the calculation of 
$\rho_{\rm vac}$ 
to keep the mass dimension 
of the cosmological constant 
fixed in Eq.~(\ref{eq3.1})
\cite{Brown:1992db}.
\footnote{
The full expression for the ZPE term in Eq.~(\ref{eq:4c}) after covariant 
dimensional regularisation is
\begin{equation}
    \rho_{\rm vac} = - p_{\rm vac} = - \hbar \ g_i \ \frac{\mu^4}{2 (4 \pi)^{(D-1)/2}} \ \frac{\Gamma(-D/2)}{\Gamma(-1/2)} \ 
    \biggl( \frac{m}{\mu} \biggr)^D
    \label{eq:5f}
\end{equation}
with Eq.~(\ref{eq4.5}) obtained in 
the limit $\epsilon \to 0$.
Eq.~(\ref{eq:5f}) respects the correct vacuum equation of state. %
The pole terms "thrown away" by the analytic dimensional continuation come with residue proportional to $\mu^4$ for the pole at  
$1/(4-\epsilon)$ 
and $m^2 \mu^2$ for the pole at $1/(2 - \epsilon)$.
}
After subtracting out the divergent pole term into a renormalisation counterterm, 
the remaining finite part 
with ${\overline {\rm MS}}$
renormalisation is 
$(\rho_{\rm zpe}
= - p_{\rm zpe})|_{\overline {\rm MS}}
=
- 
\hbar \ g_i \
\frac{m^4}{64 \pi^2} 
\bigl[ 
\frac{3}{2} 
- \ln \bigl( \frac{m^2}{4 \pi \mu^2} \bigr) \bigr]
$.

The ZPE in Eq.~(\ref{eq4.5}) 
for each given particle is 
proportional to the particle's mass to the fourth power, 
thus vanishing for 
massless photons and gluons. 
For the Standard Model particles 
(quarks, 
 charged leptons, W, Z and Higgs) 
the non vanishing ZPEs 
are generated through the BEH mechanism.
Majorana neutrinos might get their mass through the Weinberg operator
$m_\nu \sim \Lambda_{\rm ew}^2/M$ 
with mass connected to the electroweak scale $\Lambda_{\rm ew}$. 
The ZPE would vanish for 
possible massless gravitons.
For the Standard Model Higgs with mass squared 
$m_h^2 = 2 \lambda v^2$,  
the Higgs boson
ZPE develops an imaginary part if the Higgs 
self-coupling crosses zero
through the 
$m_h^4 \ln m_h^2$ term in Eq.~(\ref{eq4.5})
signalling vacuum instability for negative
$\lambda$, 
$[ \ln \lambda = \ln (- \lambda) - i \pi$ for $\lambda < 0 ]$ 
-- see Ref.~\cite{Bass:2020nrg}.

To understand the 
interpretation of the gravity term $\rho_\Lambda$ here,
the net cosmological constant is  the observable, 
not the 
individual terms 
in Eq.~(\ref{eq4.1}) 
which appear as  intermediate steps in the calculation.
The cosmological constant 
term $\rho_{\rm vac}$ is RG scale invariant so 
the RG scale dependence cancels 
with the gravity term 
$\rho_\Lambda$, 
which also restores the symmetries of the low energy vacuum below the scale of emergence 
\footnote{Note the part similarity with the Chern-Simons current 
and the axial anomaly. 
The Chern-Simons current restores gauge invariance 
in the renormalised 
axial vector current and 
also carries RG scale dependence. 
Here the $\rho_\Lambda$ 
term is restoring the 
global symmetry of spacetime translation invariance of the vacuum.
The Chern-Simons current 
in QCD involves gluon fields 
that couple to the 
quark axial-vector vertex 
beyond simple tree approximation,  
whereas here $\rho_\Lambda$ 
appears as a gravitational term in General Relativity. 
}. 
If one does wish to use the 
non-covariant 
brute force cut-off to define the ZPEs, 
then one should compensate
with a similar Lorentz violation in  
the $\rho_\Lambda$ term 
to guarantee the correct vacuum EoS for the net  $\rho_{\rm vac}$.

The role of 
$\rho_\Lambda$
here has analogy with the 
treatment of vacuum energy in quantum liquids in condensed matter physics~\cite{Volovik:2004gi}. 
The Gibbs-Duhem relation for a quantum liquid at 
zero pressure sets 
the corresponding 
vacuum energy density to zero.
One finds cancellation 
from quasiparticle 
ZPE quantum contributions 
up to the energy scale characterising 
the scale of emergence for the quantum liquid against the macroscopic 
degrees of freedom, 
e.g. atoms...,  
from above the cut-off for the low-energy theory  
(the latter playing an analogous role to $\rho_\Lambda$).
A key difference between an emergent Standard Model and these low temperature condensed matter systems is that in the condensed matter systems 
we know the degrees of freedom both above and 
below the scale of emergence whereas Planck scale physics is 
not directly accessible to experiments.

The QCD phase transition in the early Universe, 
which occurred 
when the Universe was  
about 
$10^{-5}$ seconds old, 
involves a change in the free energy 
\cite{Straumann:2004vh}
but need not change the cosmological constant
if the change in the 
quark-gluon ZPE and 
QCD potential terms is compensated by a change in the gravity term $\rho_\Lambda$ 
keeping the symmetries of the metric held fixed, at least at leading order in $1/M$. 
This would correspond 
to a change in the ultraviolet completion of the effective theory with a change in the degrees of freedom from quarks and gluons to hadrons.

We note an alternative treatment, \cite{Sola:2020} where 
a renormalisation prescription was introduced 
for eliminating the 
$m^4$ terms in the ZPEs 
equivalent to 
cancelling the ZPE contributions in curved space against the corresponding ZPE evaluated
in Minkowski space. 
This prescription when taken alone does not determine the value of the cosmological constant, which is here determined by the symmetries of the metric at different orders in the low energy expansion in powers of $1/M$.

\section{Scale hierarchies, symmetries and regularisation}
\label{sec:Scales}

In the last Section we discussed 
the appearance of different mass scales in 
the separate contributions 
to the cosmological constant
and 
the sensitivity of the ZPEs equation of state to the choice of regularisation used in evaluating it.
In this Section we extend this
discussion to a more general discussion of scale hierarchies
including the Higgs mass naturalness puzzle.

Ultraviolet regularisation is an  essential part in dealing with infinities, as a precursor to renormalisation. 
Renormalisable quantum field theories involve 
renormalisation 
counterterms parametrising 
our ignorance of deep ultraviolet physics 
(here including what is beyond the scale of emergence). 
Regularisation is needed for real life calculations in defining our notion of "infinity". 
However, regularisation 
is more than just a mathematical trick to isolate and control the ultraviolet divergences 
and comes with real physical input. 
It breaks some symmetries of the classical theory which are not necessarily all restored in the continuum~\cite{Shifman:1988zk}.
One has to be sure that the regularisation  
and final calculation 
preserves the fundamental ingredients of gauge invariance and Lorentz covariance 
(with any net violations of Lorentz invariance 
entering 
at most at ${\cal O}(\Lambda_{\rm ew}^2/M^2)$
within the emergence picture 
and classical symmetries 
preserved as much as possible.)
Beyond the issue with covariance and the ZPE equation of state 
discussed above, 
there is also the 
famous example with the 
axial anomaly 
in triangle diagrams 
with vertices 
being 
two vector currents and one axial vector current 
\cite{Adler:1969gk,Bell:1969ts}. 
Gauge invariance at the vector current vertices 
induces an anomalous term in the divergence of the singlet axial vector current which is missed in the brute force cut-off theory  \cite{Bass:1991yx}.
Among regulators, a brute force cut-off on momentum integrals breaks Lorentz invariance. 
Dimensional regularisation 
involves analytic continuation into regulator dimensions. 
Here covariance and gauge invariance are respected, 
the latter with 
suitable treatment of 
Dirac $\gamma_5$ matrices in the regulator dimensions to ensure the axial anomaly \cite{tHooft:1972tcz}. 
The connection to physical resolution is less transparent.
One typically chooses the renormalisation scale with 
${\rm \overline{MS}}$ 
to minimise logarithms 
$\log m^2/\mu^2$ in perturbative 
calculations with corresponding RG evolution in $\mu$ 
($m$ denotes a particle mass scale in the calculation). 
Lattice regularisation violates translational 
and rotational invariance symmetries in the continuum 
and gives issues with fermion doubling.
In calculations 
it is important to choose a consistent regularisation and renormalisation
procedure for all observables.

Theoretically,  
the renormalised Higgs mass squared comes with the divergent counterterm
\begin{equation}
m_{h \ {\rm bare}}^2 
= m_{h \ {\rm ren}}^2 + \delta m_h^2
\label{eq:6a}
\end{equation}
where, at leading order,  
employing a cut-off in 
the 
Higgs self-energy diagrams
give the quadratic divergence
\begin{equation}
\delta m_h^2 
=
\frac{K^2}{16 \pi^2}
\frac{6}{v^2} 
\biggl(
m_h^2 + m_Z^2 + 2 m_W^2 - 4 m_t^2
\biggr)
=
\frac{K^2}{16 \pi^2}
2 
\biggl(
\frac{9}{4} g^4 + \frac{3}{4} g'^4 + 6 \lambda - 6 y_t^2
\biggr)
\label{eq:6b}
\end{equation}
Here $K$ is an ultraviolet 
scale characterising the limit to where the Standard Model should work.
The renormalised mass is 
the mass
extracted from experiments, being 
related by a perturbative expansion 
to the particle pole mass.
For a textbook discussion
of bare and renormalised mass parameters see \cite{Pokorski:1987ed,Taylor:1976ru}.
In Eq.~(\ref{eq:6b}) 
we neglect contributions from lighter mass quarks; 
next-to-leading order 
corrections 
are suppressed by factors of $1/(4\pi)^2$.
The hierarchy or naturalness puzzle is why is $m_h^2$ 
so much smaller than the Planck mass or any other large physical scale that might characterise new physics effects beyond the Standard Model?

If one instead evaluates the Higgs mass squared using dimensional regularisation, the quadratic divergence in Eq.~(\ref{eq:6b}) corresponds to a pole at $D=2$~\cite{Veltman:1980mj}.
With 
analytic continuation in
the dimensions 
this pole is "thrown away" 
in the formal 
regularisation procedure 
and the resulting 
divergence comes out 
proportional to the 
$1/\epsilon$ 
pole term at $D=4$ 
instead of 
the large ultraviolet scale term $K^2$.  
Results are available up to two loop calculations, see \cite{Kniehl:2015nwa}.
The leading order term with ${\overline {\rm MS}}$ is 
\begin{equation}
    \delta m_h^2 
    = m_h^2 
    \frac{1}{16 \pi^2} \frac{1}{\epsilon} 
    \biggl( 
    3 y_t^2 
    + 6 \lambda 
    - \frac{9}{4} g^2 - 
    \frac{3}{4} g'^2 
    \biggr) + ...
    \label{eq:6c}
\end{equation}
Here the scale factor in $\delta m_h^2$ 
is set by the Standard Model  particle masses, 
which follows on 
dimensional grounds since these are the only mass scales in the calculation.
The different coefficients 
between Eqs.(\ref{eq:6b}) and (\ref{eq:6c}) follow since the quadratic divergence in Eq.~(\ref{eq:6b}) 
corresponds to a pole at 
$D=2$ 
whereas the coefficient in Eq.~(\ref{eq:6c}) is the pole at $D=4$
with dimensional regularisation.

If taken alone without coupling to extra particles, 
the Standard Model is 
self-consistent without any 
large scale hierarchy issue if we work in 
${\overline {\rm MS}}$ 
which involves no explicit large mass scale $K$ 
\cite{Wells:2009kq}.
Ultraviolet divergences can consistently be absorbed in renormalisation counterterms.
However, usual effective theory arguments suggest 
that, independent of this, 
mass scales 
should be close to the ultraviolet cut-off that defines the limit of the effective theory 
unless some symmetry of the Lagrangian 
like gauge invariance 
or chiral symmetry pushes them to zero. 
The Higgs mass and cosmological constant 
are not protected this way.
Hence the issue whether and what new physics one needs to suppress them. 
In the scenario discussed here it is the dynamics of the vacuum, 
its stability and spacetime symmetries with an emergent Standard Model, 
that fix the size of the Higgs mass and 
the cosmological constant with the infrared and ultraviolet connected through RG running of the Higgs self coupling $\lambda$.
Vacuum stability involves a
delicate fine tuning and conspiracy 
of Standard Model parameters.

It is interesting also to consider the 
theoretical issue whether the coefficient in Eq.~(\ref{eq:6b}) might cross zero and change sign. 
The running couplings in Eqs.~(\ref{eq:2a})-(\ref{eq:2c}) 
have different RG behaviour, see Fig.~\ref{fig:runningC}, 
and come with different signs for bosons and fermions. 
Here the result is calculation dependent. 
If $\delta m_h^2$ were to cross zero 
-- so called Veltman crossing -- 
below the Planck scale and the scale of emergence,
then this 
might be interpreted in terms of a first order phase transition with electroweak symmetry breaking
below the crossing scale.
Above this scale 
the Standard Model 
would enter a symmetric phase
with the 4 Higgs states associated with the BEH Higgs doublet field
then carrying large mass 
of order the crossing scale 
and all other Standard Model particles massless. 
This scenario was found in the calculations of 
\cite{Jegerlehner:2013cta} 
(with emergence taken at the Planck scale instead of $10^{16}$ GeV)
with a stable vacuum up to the Planck mass and 
Veltman crossing around $10^{16}$ GeV. 
If manifest in Nature, 
it would allow the Higgs 
to behave as the inflaton \cite{Jegerlehner:2014mua}.
In other calculations 
crossing was found 
not below the Planck mass in \cite{Bass:2020nrg} 
and
(with metastable vacuum)  
around $10^{20}$ GeV in \cite{Masina:2013wja}
and much above the Planck scale in 
\cite{Degrassi:2012ry} and
\cite{Hamada:2012bp}.

We close mentioning 
possible similar phenomena 
between the scale hierarchy with electroweak symmetry  breaking,  
$v \ll M_{\rm Pl}$,  
and the ferromagnetic 
phase transition in condensed matter physics. 
Below 
the phase transition the magnetisation is small 
as 
the reduced temperature
$(T-T_c)/T \to 0^-$ 
with $T_c$ 
the critical temperature 
-- that is, 
close to the phase transition. 
Whereas 
emergent gauge systems are associated with topological phase transitions, 
Higgs phenomena is associated with spontaneous symmetry breaking defined with respect to a particular gauge choice. 
Electroweak symmetry breaking 
might correspond to a Universe 
close to the phase transition and near to the critical point \cite{Jegerlehner:2013cta}.

\section{Conclusions}
\label{sec:conclusions}

The main surprise from the LHC is that the Standard Model is working so well.
Perhaps the Standard Model is more special than previously assumed.
If one assumes no coupling to other particles 
then 
RG evolution 
suggests that the Standard Model with its measured parameters 
may be working up to very high scales, 
perhaps up the scales
$10^{16}$ GeV where grand unified theories 
have previously been 
proposed to work or even up 
to the Planck scale. 
Crucial here is the RG behaviour of the Higgs 
self coupling $\lambda$ which 
is very sensitive to a 
delicate interplay of 
"low energy" Standard Model  parameters measured in laboratory experiments. 
Small changes in some of these parameters would lead to a very different theory and to very different physics. 
The infrared and ultraviolet 
limits of the Standard Model are thus strongly correlated.

If the Standard Model might really work up to these large scales, then a plausible scenario is that it might be  emergent, 
with gauge symmetries "dissolving" in the ultraviolet and particles as 
the stable long range excitations of some critical system that resides close to the Planck scale.
In this scenario, 
new physics most likely resides 
in higher dimensional operator terms that become important at energies 
very close to the scale of emergence and which may have been active in the very early Universe.
The cosmological constant comes out with a scale similar to what we expect for tiny Majorana neutrino masses. 
In this picture 
dark matter,  
if it is not primordial black holes, 
might be associated 
with axion-like particles 
or with new higher dimensional gravitational couplings 
(beyond minimal General Relativity).
Lorentz invariance might be broken at $D \geq 6$.
The question why the cosmological constant 
is so small then 
becomes the issue why Nature so likes the Minkowski metric 
with connection to a spatially flat Universe.
The Higgs mass would be environmentally selected in connection with vacuum stability, 
with 
the hierarchy puzzle a delicate conspiracy of Standard Model parameters in the low energy phase. 
It is interesting to consider possible connections between the scale of emergence and scale of inflation.

Extra CP violation beyond the 
Standard Model 
Cabibbo-Kobayashi-Maskawa matrix 
can arise through 
complex phases associated 
with Majorana neutrino masses 
and through 
new $D=6(+)$ operators. 
Baryon number violation can also occur at $D=6$
with possible 
non-equilibrium at 
the phase transition 
associated with emergence and
electroweak symmetry breaking 
providing at least qualitatively the conditions needed for baryogenesis~\cite{Sakharov:1967dj}.

The Standard Model effective theory discussed here differs
from usual discussions in particle theory \cite{Isidori:2023pyp}  where one supposes new interactions at $D=4$ at higher energies above the range of present experiments. 
The effect of these interactions is integrated out with redefinition of the low energy parameters 
as well as parametrised 
by higher dimensional terms with mass scale representing the scale of new physics, 
with new heavy particle degrees of freedom liberated when one goes through 
their production thresholds.
In our emergence scenario 
the Standard Model is taken to work at $D=4$ up to the large scale of emergence. 
The higher dimensional terms 
describe new physics from the phase transition which produces the Standard Model  but unconstrained by requirements of renormalisability at $D=4$.

With emergence open theoretical questions are the universality class for the Standard Model and any critical dimension for the topological-like phase transition that produces it.
If the Standard Model might be emergent, then what about gravitation? If the gauge symmetries of General Relativity might be emergent at a scale below the Planck mass, 
then this would alleviate the usual problems associated with quantising gravitation.

The similarity of the cosmological scale and 
values we expect for light neutrino masses is intriguing, as is the similarity between the phenomenological 
acceleration parameter 
$a_0$ and
$a_\Lambda$ involving 
the cosmological constant.
While these may be coincidences, these relations may be hinting at some deeper connection between the infrared and ultraviolet through higher dimensional terms in a low energy expansion in $1/M$ associated with emergence.
In the absence of signals of new particles one should at least treat such numerical similarities as possible clues and see where they might lead. 
Next generation experiments at the high energy, precision and cosmology frontiers will teach us much about the deeper structure of Nature. 
If the Standard Model does indeed work up to the highest scales, 
then vacuum stability 
might be pointing to deep interconnections between 
the physics of the infrared and extreme ultraviolet.

\ack
{I thank 
M. Ackermann, 
F. Jegerlehner, 
J. Krysziak 
and 
G. Volovik
for stimulating discussions 
on the physics issues discussed in this paper.
I thank the Alexander von Humboldt Foundation for support for this meeting, the Humboldt Kolleg 
"Clues to a Mysterious Universe - exploring the interface of particle, gravity and quantum physics".
}



\begin{thebibliography}{9}

  
 \bibitem{Pokorski:1987ed}
  S.~Pokorski.
  {\it Gauge Field Theories},  
2nd edition 
(Cambridge Univ. Press, 2000)


\bibitem{Taylor:1976ru}
  J.~C.~Taylor,
  {\it Gauge Theories of Weak Interactions}
(Cambridge Univ. Press, 1976)


\bibitem{Altarelli:2013tya}
G.~Altarelli,
Collider Physics within the Standard Model: a Primer,
[arXiv:1303.2842 [hep-ph]].



\bibitem{Bass:2021acr}
  S.~D.~Bass, A.~De Roeck and M.~Kado,
  The Higgs boson  implications and prospects for future discoveries,
  Nature Rev. Phys. \textbf{3} (2021) 608.
(\href{https://doi.org/10.1038/s42254-021-00341-2}{doi:10.1038/s42254-021-00341-2})


\bibitem{ATLAS:2022vkf}
 The ATLAS Collaboration,
 A detailed map of Higgs boson interactions by the ATLAS experiment ten years after the discovery,
Nature \textbf{607} (2022) no.7917, 52-59
[erratum: Nature \textbf{612} (2022) no.7941, E24].
(\href{https://doi.org/doi:10.1038/s41586-022-04893-w}
{doi:10.1038/s41586-022-04893-w})


\bibitem{CMS:2022dwd}
The CMS Collaboration,
A portrait of the Higgs boson by the CMS experiment ten years after the discovery,
Nature \textbf{607} (2022) no.7917, 60-68.
(\href{https://doi.org/doi:10.1038/s41586-022-04892-x}{doi:10.1038/s41586-022-04892-x})


\bibitem{Fan:2022eto}
X.~Fan, T.~G.~Myers, B.~A.~D.~Sukra and G.~Gabrielse,
Measurement of the Electron Magnetic Moment,
Phys. Rev. Lett. \textbf{130} (2023) no.7, 071801.
(\href{https://doi.org/10.1103/PhysRevLett.130.071801}{doi:10.1103/PhysRevLett.130.071801})


\bibitem{Roussy:2022cmp}
T.~S.~Roussy, L.~Caldwell, T.~Wright, W.~B.~Cairncross, Y.~Shagam, K.~B.~Ng, N.~Schlossberger, S.~Y.~Park, A.~Wang and J.~Ye, \textit{et al.}
A new bound on the electron's electric dipole moment,
Science \textbf{381} (2023) 46. 
(\href{https://doi.org/10.1126/science.adg4084}
{doi:10.1126/science.adg4084})


\bibitem{Jakobs:2023}
K. Jakobs and G. Zanderighi, 
The profile of the Higgs boson
- status and prospects, 
Phil. Trans. R. Soc. A \textbf{382} (2023) 20230087. 
(\href{https://doi.org/10.1098/rsta.2023.0087}
{doi:10.1098/rsta.2023.0087}).


\bibitem{LIGOScientific:2016aoc}
B.~P.~Abbott \textit{et al.} [LIGO Scientific and Virgo],
Observation of Gravitational Waves from a Binary Black Hole Merger,
Phys. Rev. Lett. \textbf{116} (2016)  061102.
(\href{https://doi.org/10.1103/PhysRevLett.116.061102}
{doi:10.1103/PhysRevLett.116.061102})


\bibitem{EventHorizonTelescope:2019dse}
K.~Akiyama \textit{et al.} [Event Horizon Telescope],
First M87 Event Horizon Telescope Results. I. The Shadow of the Supermassive Black Hole,
Astrophys. J. Lett. \textbf{875} (2019) L1.
(\href{https://doi.org/10.3847/2041-8213/ab0ec7}
{doi:10.3847/2041-8213/ab0ec7})


\bibitem{Lee:2020zjt}
J.~G.~Lee, E.~G.~Adelberger, T.~S.~Cook, S.~M.~Fleischer and B.~R.~Heckel,
New Test of the Gravitational $1/r^2$ Law at Separations down to 52 $\mu$m,
Phys. Rev. Lett. \textbf{124} (2020) 101101. 
(\href{https://doi.org/10.1103/PhysRevLett.124.101101}
{doi:10.1103/PhysRevLett.124.101101})

\bibitem{Kibble:1961ba}
T.~W.~B.~Kibble,
Lorentz invariance and the gravitational field,
J. Math. Phys. \textbf{2} (1961) 212. 
(\href{https://doi.org/10.1063/1.1703702}
{doi:10.1063/1.1703702})


\bibitem{Sciama:1964wt}
D.~W.~Sciama,
The Physical Structure of General Relativity,
Rev. Mod. Phys. \textbf{36} (1964) 463
[erratum: Rev. Mod. Phys. \textbf{36} (1964) 1103].
(\href{https://doi.org/10.1103/RevModPhys.36.1103}
{doi:10.1103/RevModPhys.36.1103})


\bibitem{Pokorski:2023}
S. Pokorski, 
After the Higgs boson discovery: a turning point in particle physics, 
Phil. Trans. R. Soc. A \textbf{382} (2023) 20230090. (\href{https://doi.org/10.1098/rsta.2023.0090}{doi:10.1098/rsta.2023.0090})


\bibitem{Dvali:2023}
G. Dvali, The role of gravity in naturalness versus consistency:
strong-$CP$ and dark energy,
Phil. Trans. R. Soc. A \textbf{382} (2023) 20230084. (\href{https://doi.org/10.1098/rsta.2023.0084}{doi:10.1098/rsta.2023.0084})


\bibitem{Heisenberg:2023}
L. Heisenberg, 
Balance Laws as Test of Gravitational Waveforms, 
Phil. Trans. R. Soc. A \textbf{382} (2023) 20230086. (\href{https://doi.org/10.1098/rsta.2023.0086}{doi:10.1098/rsta.2023.0086})


\bibitem{Krizan:2023}
P. Krizan, Flavour Physics as a Window to New Physics Searches, 
Phil. Trans. R. Soc. A \textbf{382} (2023) 20230088. (\href{https://doi.org/10.1098/rsta.2023.0088}{doi:10.1098/rsta.2023.0088})


\bibitem{Ackermann:2023}
M. Ackermann and K. Helbing, Searches for beyond-standard-model physics with
astroparticle physics instruments,
Phil. Trans. R. Soc. A \textbf{382} (2023) 20230082.
(\href{http://doi.org/10.1098/rsta.2023.0082}{doi:10.1098/rsta.2023.0082})


\bibitem{Baudis:2023}
L. Baudis, Dual-phase xenon time projection chambers for rare-event
searches, 
Phil. Trans. R. Soc. A \textbf{382} (2023) 20230083. (\href{https://doi.org/10.1098/rsta.2023.0083}{doi:10.1098/rsta.2023.0083})


\bibitem{Schieck:2023}
V. Mokina and J. Schieck,
Rare Event Searches with
Cryogenic Detectors, 
Phil. Trans. R. Soc. A \textbf{382} (2023) 20230091. (\href{https://doi.org/10.1098/rsta.2023.0091}{doi:10.1098/rsta.2023.0091})


\bibitem{Malbrunot:2023}
D. Comparat, 
C. Malbrunot, 
S. Malbrunot-Ettenauer, 
E. Widmann and P. Yzombard, 
Experimental perspectives on
the matter-antimatter
asymmetry puzzle:
developments in electron
EDM and $\overline{\rm H}$ experiments, 
Phil. Trans. R. Soc. A \textbf{382} (2023) 20230089.
(\href{https://doi.org/10.1098/rsta.2023.0089}{doi:10.1098/rsta.2023.0089})


\bibitem{Jegerlehner:2013cta}
  F.~Jegerlehner,
  The Standard model as a low-energy effective theory: what is triggering the Higgs mechanism?,
  Acta Phys.\ Polon.\ B {\bf 45} (2014)   1167.
(\href{https://doi.org/10.5506/APhysPolB.45.1167}
{doi:10.5506/APhysPolB.45.1167})


\bibitem{Bednyakov:2015sca}
A.~V.~Bednyakov, B.~A.~Kniehl, A.~F.~Pikelner and O.~L.~Veretin,
Stability of the Electroweak Vacuum: Gauge Independence and Advanced Precision,
Phys. Rev. Lett. \textbf{115} (2015) 201802.
(\href{https://doi.org/10.1103/PhysRevLett.115.201802}
{doi:10.1103/PhysRevLett.115.201802})

  
\bibitem{Bjorken:2001pe}
  J.~Bjorken,
  Emergent gauge bosons,
  hep-th/0111196. 


\bibitem{Jegerlehner:2018zxm}
F.~Jegerlehner,
The Hierarchy Problem and the Cosmological Constant Problem Revisited - A new view on the SM of particle physics,
Found. Phys. \textbf{49} (2019) no.9, 915-971. 
(\href{https://doi.org/10.1007/s10701-019-00262-2}
{doi:10.1007/s10701-019-00262-2})


\bibitem{Jegerlehner:2021vqz}
F.~Jegerlehner,
The Standard Model of Particle Physics as a Conspiracy Theory and the Possible Role of the Higgs Boson in the Evolution of the Early Universe,
Acta Phys. Polon. B \textbf{52} (2021) no.6-7, 575-605.
(\href{https://doi.org/10.5506/APhysPolB.52.575}
{doi:10.5506/APhysPolB.52.575})


\bibitem{Bass:2021wxv}
S.~D.~Bass,
Emergent gauge symmetries: making symmetry as well as breaking it,
Phil. Trans. Roy. Soc. Lond. A. 
\textbf{380} (2021) no.2216, 20210059.
(\href{https://doi.org/10.1098/rsta.2021.0059}
{doi:10.1098/rsta.2021.0059})


\bibitem{Bass:2020egf}
  S.~D.~Bass and J.~Krzysiak,
  Vacuum energy with mass generation and Higgs bosons,
  Phys.\ Lett.\ B {\bf 803} (2020) 135351.
(\href{https:doi.org/10.1016/j.physletb.2020.135351}
{doi:10.1016/j.physletb.2020.135351})


\bibitem{Bass:2020nrg}
S.~D.~Bass and J.~Krzysiak,
The cosmological constant and Higgs mass with emergent gauge symmetries,
Acta Phys. Polon. B \textbf{51} (2020) 1251.
(\href{https://doi.org/10.5506/APhysPolB.51.1251}
{doi:10.5506/APhysPolB.51.1251})


\bibitem{Higgs:1964ia}
  P.~W.~Higgs,
  Broken symmetries, massless particles and gauge fields,
  Phys.\ Lett.\  {\bf 12} (1964) 132.
(\href{https://doi.org/10.1016/0031-9163(64)91136-9}
{doi:10.1016/0031-9163(64)91136-9})


\bibitem{Higgs:1964pj}
  P.~W.~Higgs,
  Broken Symmetries and the Masses of Gauge Bosons,
  Phys.\ Rev.\ Lett.\  {\bf 13} (1964) 508.
(\href{https://doi.org/10.1103/PhysRevLett.13.508}{doi:10.1103/PhysRevLett.13.508})


\bibitem{Higgs:1966ev}
  P.~W.~Higgs,
Spontaneous Symmetry Breakdown without Massless Bosons,
  Phys.\ Rev.\  {\bf 145} (1966) 1156.
(\href{https://doi.org/10.1103/PhysRev.145.1156}{doi:10.1103/PhysRev.145.1156})


\bibitem{Englert:1964et}
  F.~Englert and R.~Brout,
Broken Symmetry and the Mass of Gauge Vector Mesons,
  Phys.\ Rev.\ Lett.\  {\bf 13} (1964) 321.
(\href{https://doi.org/10.1103/PhysRevLett.13.321}
{doi:10.1103/PhysRevLett.13.321})


\bibitem{Veltman:1997nm}
  M.~J.~G.~Veltman,
  {\it Reflections on the Higgs system},
  CERN-97-05, CERN-YELLOW-97-05.
(\href{https://doi.org/10.5170/CERN-1997-005}
{doi:10.5170/CERN-1997-005})
  

\bibitem{Kibble:2014gug}
  T.~W.~B.~Kibble,
  Spontaneous symmetry breaking in gauge theories,
  Phil.\ Trans.\ Roy.\ Soc.\ Lond.\ A {\bf 373} (2014) 20140033.
(\href{https://doi.org/10.1098/rsta.2014.0033}
{doi:10.1098/rsta.2014.0033})
  
  
\bibitem{LlewellynSmith:1973yud}
  C.~H.~Llewellyn Smith,
  High-Energy Behavior and Gauge Symmetry,
  Phys.\ Lett.\  B {\bf 46} (1973) 233.
(\href{https://doi.org/10.1016/0370-2693(73)90692-8}
{doi:10.1016/0370-2693(73)90692-8})  


\bibitem{Bell:1973ex}
  J.~S.~Bell,
  High-energy Behavior Of Tree Diagrams In Gauge Theories,
  Nucl.\ Phys.\ B {\bf 60} (1973) 427.
(\href{https://doi.org/10.1016/0550-3213(73)90191-0}
{doi:10.1016/0550-3213(73)90191-0})

  
\bibitem{Cornwall:1973tb}
  J.~M.~Cornwall, D.~N.~Levin and G.~Tiktopoulos,
  Uniqueness of spontaneously broken gauge theories,
  Phys.\ Rev.\ Lett.\  {\bf 30} (1973) 1268
   Erratum: [Phys.\ Rev.\ Lett.\  {\bf 31} (1973) 572].
(\href{https://doi.org/10.1103/PhysRevLett.31.572}
{doi:10.1103/PhysRevLett.31.572}, 1
\href{https://doi.org/10.1103/PhysRevLett.30.1268}
{doi:10.1103/PhysRevLett.30.1268})  

  
\bibitem{Cornwall:1974km}
  J.~M.~Cornwall, D.~N.~Levin and G.~Tiktopoulos,
  Derivation of Gauge Invariance from High-Energy Unitarity Bounds on the S Matrix,
  Phys.\ Rev.\ D {\bf 10} (1974) 1145
   Erratum: [Phys.\ Rev.\ D {\bf 11} (1975) 972].
 (\href{https://doi.org/10.1103/PhysRevD.10.1145} 
{doi:10.1103/PhysRevD.10.1145},
\href{https://doi.org/10.1103/PhysRevD.11.972}
{doi:10.1103/PhysRevD.11.972})  
 
\bibitem{tHooft:1971qjg}
  G.~'t Hooft,
  Renormalizable Lagrangians for Massive Yang-Mills Fields,
  Nucl.\ Phys.\ B {\bf 35} (1971) 167.
(\href{https://doi.org/10.1016/0550-3213(71)90139-8}
{doi:10.1016/0550-3213(71)90139-8})

  
\bibitem{tHooft:1972tcz}
  G.~'t Hooft and M.~J.~G.~Veltman,
  Regularization and Renormalization of Gauge Fields,
  Nucl.\ Phys.\ B {\bf 44} (1972) 189.
(\href{https://doi.org/10.1016/0550-3213(72)90279-9}
{doi:10.1016/0550-3213(72)90279-9}) 

  
\bibitem{Veltman:1968ki}
  M.~J.~G.~Veltman,
  Perturbation theory of massive Yang-Mills fields,
  Nucl.\ Phys.\ B {\bf 7} (1968) 637.
(\href{https://doi.org/10.1016/0550-3213(68)90197-1}
{doi:10.1016/0550-3213(68)90197-1})  


\bibitem{Degrassi:2012ry}
G.~Degrassi, S.~Di Vita, J.~Elias-Miro, J.~R.~Espinosa, G.~F.~Giudice, G.~Isidori and A.~Strumia,
Higgs mass and vacuum stability in the Standard Model at NNLO,
JHEP \textbf{08} (2012) 098.
(\href{https://doi.org/10.1007/JHEP08(2012)098}
{doi:10.1007/JHEP08(2012)098})


\bibitem{Buttazzo:2013uya}
D.~Buttazzo, G.~Degrassi, P.~P.~Giardino, G.~F.~Giudice, F.~Sala, A.~Salvio and A.~Strumia,
Investigating the near-criticality of the Higgs boson,
JHEP \textbf{12} (2013) 089. 
(\href{https://doi.org/10.1007/JHEP12(2013)089}
{doi:10.1007/JHEP12(2013)089})


\bibitem{Kniehl:2016enc}
  B.~A.~Kniehl, A.~F.~Pikelner and O.~L.~Veretin,
  mr: a C++ library for the matching and running of the Standard Model parameters,
  Comput.\ Phys.\ Commun.\ {\bf 206} (2016) 84.
(\href{https://doi.org/10.1016/j.cpc.2016.04.017}
{doi:10.1016/j.cpc.2016.04.017})


\bibitem{Hambye:1996wb}
T.~Hambye and K.~Riesselmann,
Matching conditions and Higgs mass upper bounds revisited,
Phys. Rev. D \textbf{55} (1997) 7255-7262. 
(\href{https://doi.org/10.1103/PhysRevD.55.7255}
{doi:10.1103/PhysRevD.55.7255})


\bibitem{CMS:2023ebf}
The CMS Collaboration,
Measurement of the top quark mass using a profile likelihood approach with the lepton+jets final states in proton-proton collisions at $\sqrt{s}$ = 13 TeV,
Eur. Phys. J. C \textbf{83} (2023) 963.
(\href{https://doi.org/10.1140/epjc/s10052-023-12050-4}{doi:10.1140/epjc/s10052-023-12050-4})


\bibitem{Hoang:2020iah}
A.~H.~Hoang,
What is the Top Quark Mass?
Ann. Rev. Nucl. Part. Sci. \textbf{70} (2020) 225-255.
(\href{https://doi.org/10.1146/annurev-nucl-101918-023530}
{doi:10.1146/annurev-nucl-101918-023530})


\bibitem{Branchina:2013jra}
V.~Branchina and E.~Messina,
Stability, Higgs Boson Mass and New Physics,
Phys. Rev. Lett. \textbf{111} (2013), 241801.
(\href{https://doi.org/10.1103/PhysRevLett.111.241801}
{doi:10.1103/PhysRevLett.111.241801})


\bibitem{Anderson:1972pca}
  P.~W.~Anderson,
  More Is Different,
  Science {\bf 177} (1972) 393.
(\href{https://doi.org/10.1126/science.177.4047.393}
{doi:10.1126/science.177.4047.393})


\bibitem{Palacios:book}
P. Palacios,
{\it Emergence and Reduction in Physics} 
(Cambridge Univ. Press, 2022).


\bibitem{Baskaran:1987my}
  G.~Baskaran and P.~W.~Anderson,
  Gauge theory of high temperature superconductors and strongly correlated Fermi systems,
  Phys.\ Rev.\ B {\bf 37} (1988) 580.
(\href{https://doi.org/10.1103/PhysRevB.37.580}
{doi:10.1103/PhysRevB.37.580})


\bibitem{Affleck:1988zz}
  I.~Affleck, Z.~Zou, T.~Hsu and P.~W.~Anderson,
  SU(2) gauge symmetry of the large-U limit of the Hubbard model,
  Phys.\ Rev.\ B {\bf 38} (1988) 745.
(\href{https://doi.org/10.1103/PhysRevB.38.745}
{doi:10.1103/PhysRevB.38.745})


\bibitem{Volovik:2003fe}
  G.~E.~Volovik,
  {\it The Universe in a helium droplet},
  Int.\ Ser.\ Monogr.\ Phys.\  {\bf 117} (2006) 1
  %
  (Oxford Univ. Press, 2006).

 
\bibitem{Volovik:2008dd}
G.~E.~Volovik,
Emergent physics: Fermi point scenario,
Phil. Trans. Roy. Soc. Lond. A \textbf{366} (2008) 2935.
(\href{https://doi.org/10.1098/rsta.2008.0070}
{doi:10.1098/rsta.2008.0070})


\bibitem{Sachdev:2015slk}
  S.~Sachdev,
  Emergent gauge fields and the high temperature superconductors,
  Phil.\ Trans.\ Roy.\ Soc.\ Lond.\ A {\bf 374} (2016) 20150248.
(\href{https://doi.org/10.1098/rsta.2015.0248}
{doi:10.1098/rsta.2015.0248})


\bibitem{Levin:2004js}
  M.~A.~Levin and X.~G.~Wen,
  Colloquium: Photons and electrons as emergent phenomena,
  Rev.\ Mod.\ Phys.\  {\bf 77} (2005) 871.
(\href{https://doi.org/10.1103/RevModPhys.77.871}
{doi:10.1103/RevModPhys.77.871})


\bibitem{Zaanen:2011hm}
J.~Zaanen and A.~J.~Beekman,
The Emergence of gauge invariance: The Stay-at-home gauge versus local-global duality,
Annals Phys. \textbf{327} (2012) 1146.
(\href{https://doi.org/10.1016/j.aop.2011.11.006}
{doi:10.1016/j.aop.2011.11.006})


\bibitem{Wilson:1973jj}
  K.~G.~Wilson and J.~B.~Kogut,
  The Renormalization group and the epsilon expansion,
  Phys.\ Rept.\  {\bf 12} (1974) 75.
(\href{https://doi.org/10.1016/0370-1573(74)90023-4}
{doi:10.1016/0370-1573(74)90023-4})
  
  
\bibitem{Peskin:1995ev}
  M.~E.~Peskin and D.~V.~Schroeder,
  {\it An introduction to quantum field theory}
	(Westview Press, 1995). 


\bibitem{Jegerlehner:1978nk}
  F.~Jegerlehner,
  The Vector Boson and Graviton Propagators in the Presence of Multipole Forces,
  Helv.\ Phys.\ Acta {\bf 51} (1978) 783. 


\bibitem{Forster:1980dg}
  D.~Forster, H.~B.~Nielsen and M.~Ninomiya,
  Dynamical Stability of Local Gauge Symmetry: Creation of Light from Chaos,
  Phys.\ Lett.\  B {\bf 94} (1980) 135.
(\href{https://doi.org/10.1016/0370-2693(80)90842-4}
{doi:10.1016/0370-2693(80)90842-4})


\bibitem{Jegerlehner:1998kt}
  F.~Jegerlehner,
  The 'Ether world' and elementary particles,
  hep-th/9803021.


\bibitem{tHooft:2007nis}
G.~'t Hooft,
Emergent Quantum Mechanics and Emergent Symmetries,
AIP Conf. Proc. \textbf{957} (2007) no.1, 154-163.
(\href{https://doi.org/10.1063/1.2823751}
{doi:10.1063/1.2823751})


\bibitem{Weinberg:2018apv}
S.~Weinberg,
Essay: Half a Century of the Standard Model,
Phys. Rev. Lett. \textbf{121} (2018) no.22, 220001.
(\href{https://doi.org/10.1103/PhysRevLett.121.220001}
{doi:10.1103/PhysRevLett.121.220001})


\bibitem{Bass:2020gpp}
  S.~D.~Bass,
  Emergent Gauge Symmetries and Particle Physics,
  Prog.\ Part.\ Nucl.\ Phys.\  {\bf 113} (2020) 103756.
(\href{https://doi.org/10.1016/j.ppnp.2020.103756}
{doi:10.1016/j.ppnp.2020.103756})


\bibitem{Witten:2017hdv}
  E.~Witten,
  Symmetry and emergence,
  Nature Phys.\ {\bf 14} (2018) 116.
(\href{https://doi.org/10.1038/nphys4348}
{doi:10.1038/nphys4348})

  
\bibitem{Weinberg:1979sa}
  S.~Weinberg,
  Baryon and lepton nonconserving processes,
  Phys.\ Rev.\ Lett.\ {\bf 43} (1979) 1566.
(\href{https://doi.org/10.1103/PhysRevLett.43.1566}
{doi:10.1103/PhysRevLett.43.1566})


\bibitem{Wilczek:1979hc}
  F.~Wilczek and A.~Zee,
  Operator Analysis of Nucleon Decay,
  Phys.\ Rev.\ Lett.\  {\bf 43} (1979) 1571.
(\href{https://doi.org/10.1103/PhysRevLett.43.1571}{
doi:10.1103/PhysRevLett.43.1571})


\bibitem{Grzadkowski:2010es}
B.~Grzadkowski, M.~Iskrzynski, M.~Misiak and J.~Rosiek,
Dimension-Six Terms in the Standard Model Lagrangian,
JHEP \textbf{10} (2010) 085.
(\href{https://doi.org/10.1007/JHEP10(2010)085}
{doi:10.1007/JHEP10(2010)085})


\bibitem{Slade:2019bjo}
E.~Slade,
Towards global fits in EFT's and New Physics implications,
PoS \textbf{LHCP2019} (2019) 150.
(\href{https://doi.org/10.22323/1.350.0150}
{doi:10.22323/1.350.0150})


\bibitem{Ellis:2020unq}
J.~Ellis, M.~Madigan, K.~Mimasu, V.~Sanz and T.~You,
Top, Higgs, Diboson and Electroweak Fit to the Standard Model Effective Field Theory,
JHEP \textbf{04} (2021)  279. 
(\href{https://doi.org/10.1007/JHEP04(2021)279}{doi:10.1007/JHEP04(2021)279})


\bibitem{Wetterich:2016qee}
  C.~Wetterich,
  Gauge symmetry from decoupling,
  Nucl.\ Phys.\ B {\bf 915} (2017) 135.
(\href{https://doi.org/10.1016/j.nuclphysb.2016.12.008}
{doi:10.1016/j.nuclphysb.2016.12.008})


\bibitem{Bjorken:1963vg}
  J.~D.~Bjorken,
  A Dynamical origin for the electromagnetic field,
  Annals Phys.\  {\bf 24} (1963) 174.
(\href{https://doi.org/10.1016/0003-4916(63)90069-1}
{doi:10.1016/0003-4916(63)90069-1})

   
\bibitem{Bjorken:2010qx}
  J.~D.~Bjorken,
  Emergent Photons and Gravitons: The Problem of Vacuum Structure,
  arXiv:1008.0033 [hep-ph].


\bibitem{Chkareuli:2001xe}
  J.~L.~Chkareuli, C.~D.~Froggatt and H.~B.~Nielsen,
  Lorentz invariance and origin of symmetries,
  Phys.\ Rev.\ Lett.\  {\bf 87} (2001) 091601.
(\href{https://doi.org/10.1103/PhysRevLett.87.091601}
{doi:10.1103/PhysRevLett.87.091601})


\bibitem{Weinberg:1988cp}
  S.~Weinberg,
 The Cosmological Constant Problem,
  Rev.\ Mod.\ Phys.\  {\bf 61} (1989) 1.
(\href{https://doi.org/10.1103/RevModPhys.61.1}
{doi:10.1103/RevModPhys.61.1})


\bibitem{Planck:2018vyg}
N.~Aghanim \textit{et al.} [Planck Collaboration],
Planck 2018 results. VI. Cosmological parameters,
Astron. Astrophys. \textbf{641} (2020), A6
[erratum: Astron. Astrophys. \textbf{652} (2021), C4].
(\href{https://doi.org/10.1051/0004-6361/201833910}
{doi:10.1051/0004-6361/201833910})

\bibitem{Escamilla:2023oce}
L.~A.~Escamilla, W.~Giar\`e, E.~Di Valentino, R.~C.~Nunes and S.~Vagnozzi,
The state of the dark energy equation of state circa 2023,
[arXiv:2307.14802 [astro-ph.CO]].



\bibitem{Pauli:1933b}
W.~Pauli, 
{\it Die allgemeinen Prinzipien der Wellenmechanik}, 
Handbuch
der Physik, Vol. XXIV (1933). 
New edition by N. Straumann, Springer Berlin Heidelberg 
(1990); see Appendix III, p. 202.
(\href{https://doi.org/10.1007/978-3-642-61287-9}
{doi:10.1007/978-3-642-61287-9})


\bibitem{Zeldovich:1967gd}
Y.~B.~Zel'dovich,
Cosmological Constant and Elementary Particles,
JETP Lett. \textbf{6} (1967) 316.


\bibitem{Dreitlein:1974sa}
J.~Dreitlein,
Broken symmetry and the cosmological constant,
Phys. Rev. Lett. \textbf{33} (1974), 1243-1244.
(\href{https://doi.org/10.1103/PhysRevLett.33.1243}
{doi:10.1103/PhysRevLett.33.1243})

\bibitem{Straumann:2002tv}
N.~Straumann,
The History of the cosmological constant problem,
[arXiv:gr-qc/0208027 [gr-qc]].



\bibitem{Kragh:2014jaa}
H.~S.~Kragh and J.~M.~Overduin,
The weight of the vacuum: A scientific history of dark energy,
Springer, 2014,
ISBN 978-3-642-55089-8, 978-3-642-55090-4.
(\href{https://doi.org/10.1007/978-3-642-55090-4}
{doi:10.1007/978-3-642-55090-4})


\bibitem{Bjorken:2001yv}
J.~D.~Bjorken,
Standard model parameters and the cosmological constant,
Phys. Rev. D \textbf{64} (2001) 085008.
(\href{https://doi.org/10.1103/PhysRevD.64.085008}
{doi:10.1103/PhysRevD.64.085008})


\bibitem{Altarelli:2004cp}
  G.~Altarelli,
  Neutrino 2004: Concluding talk,
  Nucl.\ Phys.\ Proc.\ Suppl.\  {\bf 143} (2005) 470.
(\href{https://doi.org/10.1016/j.nuclphysbps.2005.01.146}
{doi:10.1016/j.nuclphysbps.2005.01.146})


\bibitem{BahaBalantekin:2018ppj}
A.~B. Balantekin and B.~Kayser,
On the Properties of Neutrinos,
Ann. Rev. Nucl. Part. Sci. \textbf{68} (2018) 313. 
(\href{https://doi.org/10.1146/annurev-nucl-101916-123044}
{doi:10.1146/annurev-nucl-101916-123044})


\bibitem{Dvali:2020etd}
G.~Dvali,
$S$-Matrix and Anomaly of de Sitter,
Symmetry \textbf{13} (2020) no.1, 3. 
(\href{https://doi.org/10.3390/sym13010003}
{doi:10.3390/sym13010003})


\bibitem{Dvali:2017eba}
G.~Dvali, C.~Gomez and S.~Zell,
Quantum Break-Time of de Sitter,
JCAP \textbf{06} (2017), 028.
(\href{https://doi.org/10.1088/1475-7516/2017/06/028}
{doi:10.1088/1475-7516/2017/06/028})


\bibitem{Berezhiani:2021zst}
L.~Berezhiani, G.~Dvali and O.~Sakhelashvili,
de Sitter space as a BRST invariant coherent state of gravitons,
Phys. Rev. D \textbf{105} (2022) no.2, 025022. 
(\href{https://doi.org/10.1103/PhysRevD.105.025022}
{doi:10.1103/PhysRevD.105.025022})


\bibitem{Wetterich:2007kr}
C.~Wetterich,
Growing neutrinos and cosmological selection,
Phys. Lett. B \textbf{655} (2007) 201.
(\href{https://doi.org/10.1016/j.physletb.2007.08.060}
{doi:10.1016/j.physletb.2007.08.060})


\bibitem{Brookfield:2005bz}
A.~W.~Brookfield, C.~van de Bruck, D.~F.~Mota and D.~Tocchini-Valentini,
Cosmology of mass-varying neutrinos driven by quintessence: theory and observations,
Phys. Rev. D \textbf{73} (2006) 083515
[erratum: Phys. Rev. D \textbf{76} (2007), 049901].
(\href{https://doi.org/10.1103/PhysRevD.73.083515}
{doi:10.1103/PhysRevD.73.083515})

\bibitem{Fardon:2003eh}
R.~Fardon, A.~E.~Nelson and N.~Weiner,
Dark energy from mass varying neutrinos,
JCAP \textbf{10} (2004)  005.
(\href{https://doi.org/10.1088/1475-7516/2004/10/005}
{doi:10.1088/1475-7516/2004/10/005})


\bibitem{Riess:2019qba}
A.~G.~Riess,
The Expansion of the Universe is Faster than Expected,
Nature Rev. Phys. \textbf{2} (2019) no.1, 10-12.
(\href{https://doi.org/10.1038/s42254-019-0137-0}
{doi:10.1038/s42254-019-0137-0})

\bibitem{Efstathiou:1990xe}
G.~Efstathiou, W.~J.~Sutherland and S.~J.~Maddox,
The cosmological constant and cold dark matter,
Nature \textbf{348} (1990), 705-707.
(\href{https://doi.org/10.1038/348705a0}
{doi:10.1038/348705a0})


\bibitem{SupernovaSearchTeam:1998fmf}
A.~G.~Riess \textit{et al.} [Supernova Search Team],
Observational evidence from supernovae for an accelerating universe and a cosmological constant,
Astron. J. \textbf{116} (1998) 1009-1038.
(\href{https://doi.org/10.1086/300499}
{doi:10.1086/300499})

\bibitem{SupernovaCosmologyProject:1998vns}
S.~Perlmutter \textit{et al.} [Supernova Cosmology Project],
Measurements of $\Omega$ and $\Lambda$ from 42 high redshift supernovae,
Astrophys. J. \textbf{517} (1999) 565-586.
(\href{https://doi.org/10.1086/307221}
{doi:10.1086/307221})

\bibitem{Baumann:2008bn}
D.~Baumann and H.~V.~Peiris,
Cosmological Inflation: Theory and Observations,
Adv. Sci. Lett. \textbf{2} (2009), 105-120.
(\href{https://doi.org/10.1166/asl.2009.1019}
{doi:10.1166/asl.2009.1019})


\bibitem{Ellis:PDG}
J.~Ellis and D.~Wands, 
Inflation,
Chapter 23 of 
{\it The Review of Particle Physics (2023)} 
[Particle Data Group],
PTEP \textbf{2022} (2022), 083C01.
(\href{https://doi.org/10.1093/ptep/ptac097}
{doi:10.1093/ptep/ptac097})

\bibitem{Weinberg:1987dv}
S.~Weinberg,
Anthropic Bound on the Cosmological Constant,
Phys. Rev. Lett. \textbf{59} (1987) 2607.
(\href{https://doi.org/10.1103/PhysRevLett.59.2607}
{doi:10.1103/PhysRevLett.59.2607})

\bibitem{Baudis:2018bvr}
  L.~Baudis,
  The Search for Dark Matter,
  European Review {\bf 26} (2018) 70. (\href{https://doi.org/10.1017/S1062798717000783}
{doi:10.1017/S1062798717000783}) 


\bibitem{Bertone:2018krk}
G.~Bertone and T.~M~.P.~Tait,
A new era in the search for dark matter,
Nature \textbf{562} (2018) no.7725, 51-56.
(\href{https://doi.org/10.1038/s41586-018-0542-z}
{doi:10.1038/s41586-018-0542-z})


\bibitem{Wechsler:2018pic}
R.~H.~Wechsler and J.~L.~Tinker,
The Connection between Galaxies and their Dark Matter Halos,
Ann. Rev. Astron. Astrophys. \textbf{56} (2018) 435-487.
(\href{https://doi.org/10.1146/annurev-astro-081817-051756}
{doi:10.1146/annurev-astro-081817-051756})


\bibitem{Verlinde:2016toy}
E.~P.~Verlinde,
Emergent Gravity and the Dark Universe, 
SciPost Phys. \textbf{2} (2017) no.3, 016.
(\href{https://doi.org/10.21468/SciPostPhys.2.3.016}
{doi:10.21468/SciPostPhys.2.3.016}) 


\bibitem{Kawasaki:2013ae}
M.~Kawasaki and K.~Nakayama,
Axions: Theory and Cosmological Role,
Ann. Rev. Nucl. Part. Sci. \textbf{63} (2013)  69-95.
(\href{https://doi.org/10.1146/annurev-nucl-102212-170536}
{doi:10.1146/annurev-nucl-102212-170536})


\bibitem{Peccei:1977hh}
R.~D.~Peccei and H.~R.~Quinn,
CP Conservation in the Presence of Instantons,
Phys. Rev. Lett. \textbf{38} (1977), 1440-1443.
(\href{https://doi.org/10.1103/PhysRevLett.38.1440}
{doi:10.1103/PhysRevLett.38.1440})


\bibitem{Weinberg:1977ma}
S.~Weinberg,
A New Light Boson?,
Phys. Rev. Lett. \textbf{40} (1978) 223.
(\href{https://doi.org/10.1103/PhysRevLett.40.223}
{doi:10.1103/PhysRevLett.40.223})


\bibitem{Wilczek:1977pj}
F.~Wilczek,
Problem of Strong  $P$  and  $T$  Invariance in the Presence of Instantons,
Phys. Rev. Lett. \textbf{40} (1978) 279.
(\href{https://doi.org/10.1103/PhysRevLett.40.279}
{doi:10.1103/PhysRevLett.40.279})


\bibitem{Nakamura:2021meh}
Y.~Nakamura and G.~Schierholz,
The strong CP problem solved by itself due to long-distance vacuum effects,
Nucl. Phys. B \textbf{986} (2023)  116063. 
(\href{https://doi.org/10.1016/j.nuclphysb.2022.116063}
{doi:10.1016/j.nuclphysb.2022.116063})

\bibitem{Dvali:1995ce}
G.~R.~Dvali,
Removing the cosmological bound on the axion scale,
[arXiv:hep-ph/9505253 [hep-ph]].


\bibitem{Sikivie:2009qn}
P.~Sikivie and Q.~Yang,
Bose-Einstein Condensation of Dark Matter Axions,
Phys. Rev. Lett. \textbf{103} (2009) 111301.
(\href{https://doi.org/10.1103/PhysRevLett.103.111301}
{doi:10.1103/PhysRevLett.103.111301})


\bibitem{Carr:2021bzv}
B.~Carr and F.~Kuhnel,
Primordial black holes as dark matter candidates,
SciPost Phys. Lect. Notes \textbf{48} (2022) 1.
(\href{https://doi.org/10.21468/SciPostPhysLectNotes.48}
{doi:10.21468/SciPostPhysLectNotes.48})


\bibitem{Green:2020jor}
A.~M.~Green and B.~J.~Kavanagh,
Primordial Black Holes as a dark matter candidate,
J. Phys. G \textbf{48} (2021) no.4, 043001.
(\href{https://doi.org/10.1088/1361-6471/abc534}
{doi:10.1088/1361-6471/abc534})


\bibitem{Boehm:2020jwd}
C.~Boehm, A.~Kobakhidze, C.~A.~J.~O'hare, Z.~S.~C.~Picker and M.~Sakellariadou,
Eliminating the LIGO bounds on primordial black hole dark matter,
JCAP \textbf{03} (2021)  078.
(\href{https://doi.org/10.1088/1475-7516/2021/03/078}
{doi:10.1088/1475-7516/2021/03/078})


\bibitem{Bertone:2019irm}
G.~Bertone, D.~Croon, M.~A.~Amin, K.~K.~Boddy, B.~J.~Kavanagh, K.~J.~Mack, P.~Natarajan, T.~Opferkuch, K.~Schutz and V.~Takhistov, \textit{et al.}
Gravitational wave probes of dark matter: challenges and opportunities,
SciPost Phys. Core \textbf{3} (2020) 007. 
(\href{https://doi.org/10.21468/SciPostPhysCore.3.2.007}
{doi:10.21468/SciPostPhysCore.3.2.007})


\bibitem{Ferrarese:2002ct}
L.~Ferrarese,
Beyond the bulge: a fundamental relation between supermassive black holes and dark matter halos,
Astrophys. J. \textbf{578} (2002) 90-97.
(\href{https://doi.org/10.1086/342308}
{doi:10.1086/342308})


\bibitem{Ferrarese:2004qr}
L.~Ferrarese and H.~Ford,
Supermassive black holes in galactic nuclei: Past, present and future research,
Space Sci. Rev. \textbf{116} (2005) 523.
(\href{https://doi/org/10.1007/s11214-005-3947-6}
{doi:10.1007/s11214-005-3947-6})

 
\bibitem{McGaugh:2000sr}
  S.~S.~McGaugh, J.~M.~Schombert, G.~D.~Bothun and W.~J.~G.~de Blok,
  The Baryonic Tully-Fisher relation,
  Astrophys.\ J.\ Lett.\  {\bf 533} (2000) L99.
  (\href{https://doi.org/10.1086/312628}
{doi:10.1086/312628})
  

\bibitem{Milgrom:2020cch}
M.~Milgrom,
The $a_0$ -- cosmology connection in MOND,
[arXiv:2001.09729 [astro-ph.GA]].


\bibitem{Milgrom:1983pn}
  M.~Milgrom,
  A Modification of the Newtonian dynamics: Implications for galaxies,
  Astrophys.\ J.\  {\bf 270} (1983) 371.
(\href{https://doi.org/10.1086/161131}{doi:10.1086/161131})


\bibitem{Mayer:2022qhk}
A.~C.~Mayer, A.~F.~Teklu, K.~Dolag and R.~S.~Remus,
\ensuremath{\Lambda}CDM with baryons versus MOND: The time evolution of the universal acceleration scale in the Magneticum simulations,
Mon. Not. Roy. Astron. Soc. \textbf{518} (2022) no.1, 257-269.
(\href{https://doi.org/10.1093/mnras/stac3017}{
doi:10.1093/mnras/stac3017})


\bibitem{LIGOScientific:2017zic}
B.~P.~Abbott \textit{et al.} [LIGO Scientific, Virgo, Fermi-GBM and INTEGRAL],
Gravitational Waves and Gamma-rays from a Binary Neutron Star Merger: GW170817 and GRB 170817A,
Astrophys. J. Lett. \textbf{848} (2017) no.2, L13. 
(\href{https://doi.org/10.3847/2041-8213/aa920c}{
doi:10.3847/2041-8213/aa920c})


\bibitem{IceCube:2018dnn}
M.~G.~Aartsen \textit{et al.} [IceCube, Fermi-LAT, MAGIC, AGILE, ASAS-SN, HAWC, H.E.S.S., INTEGRAL, Kanata, Kiso, Kapteyn, Liverpool Telescope, Subaru, Swift NuSTAR, VERITAS and VLA/17B-403],
Multimessenger observations of a flaring blazar coincident with high-energy neutrino IceCube-170922A,
Science \textbf{361} (2018) no.6398, eaat1378. 
(\href{https://doi.org/10.1126/science.aat1378}
{doi:10.1126/science.aat1378})


\bibitem{Kun:2023uld}
E.~Kun, I.~Bartos, J.~Becker Tjus, P.~L.~Biermann, A.~Franckowiak, F.~Halzen and G.~Mez\H{o},
Searching for temporary gamma-ray dark blazars associated with IceCube neutrinos,
Astron. Astrophys. \textbf{679} (2023), A46.
(\href{https://doi.org/doi:10.1051/0004-6361/202346710}{
doi:10.1051/0004-6361/202346710})


\bibitem{Charlton:2020kie}
M.~Charlton, S.~Eriksson and G.~M.~Shore,
{\it
Antihydrogen and Fundamental Physics}
(Springer, 2020).
(\href{https://doi.org/10.1007/978-3-030-51713-7}{
doi:10.1007/978-3-030-51713-7})


\bibitem{Shore:2004sh}
  G.~M.~Shore,
  Strong equivalence, Lorentz and CPT violation, anti-hydrogen spectroscopy and gamma-ray burst polarimetry,
  Nucl.\ Phys.\ B {\bf 717} (2005) 86.
(\href{https://doi.org/10.1016/j.nuclphysb.2005.03.040}{doi:10.1016/j.nuclphysb.2005.03.040})


\bibitem{Jaffe:2005vp}
R.~L.~Jaffe,
The Casimir effect and the quantum vacuum,
Phys. Rev. D \textbf{72} (2005) 021301.
(\href{https://doi.org/10.1103/PhysRevD.72.021301}
{doi:10.1103/PhysRevD.72.021301})


\bibitem{Sola:2013gha}
  J.~Sola,
  Cosmological constant and vacuum energy: old and new ideas,
  J.\ Phys.\ Conf.\ Ser.\  {\bf 453} (2013) 012015.
(\href{https://doi.org/10.1088/1742-6596/453/1/012015}
{doi:10.1088/1742-6596/453/1/012015})


\bibitem{Martin:2012bt}
  J.~Martin,
  Everything you always wanted to know about the cosmological constant problem (but were afraid to ask),
  Comptes Rendus Physique {\bf 13} (2012) 566.
(\href{https://doi.org/10.1016/j.crhy.2012.04.008}
{doi:10.1016/j.crhy.2012.04.008})



\bibitem{Peebles:2002gy}
P.~J.~E.~Peebles and B.~Ratra,
The Cosmological Constant and Dark Energy,
Rev. Mod. Phys. \textbf{75} (2003), 559-606.
(\href{https://doi.org/10.1103/RevModPhys.75.559}
{doi:10.1103/RevModPhys.75.559})


\bibitem{Brown:1992db}
L.~S.~Brown,
{\it Quantum field theory},
Cambridge University Press, 1994,
ISBN 978-0-521-46946-3


\bibitem{Volovik:2004gi}
G.~E.~Volovik,
Cosmological constant and vacuum energy,
Annalen Phys. \textbf{14} (2005) 165.
(\href{https://doi.org/10.1002/andp.200410123}
{doi:10.1002/andp.200410123})


\bibitem{Straumann:2004vh}
N.~Straumann,
Cosmological phase transitions,
[arXiv:astro-ph/0409042 [astro-ph]].


\bibitem{Sola:2020}
C.~Moreno-Pulido and J.~Sola,
Running vacuum in quantum field theory in curved spacetime: renormalizing $\rho_{vac}$ without $\sim m^4$ terms,
Eur. Phys. J. C \textbf{80} (2020) no.8, 692.
(\href{https://doi.org/10.1140/epjc/s10052-020-8238-6}
{doi:10.1140/epjc/s10052-020-8238-6})


\bibitem{Shifman:1988zk}
M.~A.~Shifman,
Anomalies in gauge theories,
Phys. Rept.
\textbf{209} (1991) 341.
(\href{https://doi.org/10.1016/0370-1573(91)90020-M}
{doi.org/10.1016/0370-1573(91)90020-M})


\bibitem{Adler:1969gk}
  S.~L.~Adler,
  Axial vector vertex in spinor electrodynamics,
  Phys.\ Rev.\  {\bf 177} (1969) 2426.
(\href{https://doi.org/10.1103/PhysRev.177.2426}
{doi:10.1103/PhysRev.177.2426})


\bibitem{Bell:1969ts}
  J.~S.~Bell and R.~Jackiw,
  A PCAC puzzle: $\pi^0 \to \gamma \gamma$ in the $\sigma$ model,
  Nuovo Cim.\ A {\bf 60} (1969) 47.
(\href{https://doi.org/10.1007/BF02823296}
{doi:10.1007/BF02823296})


\bibitem{Bass:1991yx}
S.~D.~Bass, B.~L.~Ioffe, N.~N.~Nikolaev and A.~W.~Thomas,
On the infrared contribution to the photon - gluon scattering and the proton spin content,
J. Moscow. Phys. Soc. \textbf{1} (1991), 317-333.


\bibitem{Veltman:1980mj}
  M.~J.~G.~Veltman,
  The infrared-ultraviolet connection,
  Acta Phys.\ Polon.\ B {\bf 12} (1981) 437.


\bibitem{Kniehl:2015nwa}
B.~A.~Kniehl, A.~F.~Pikelner and O.~L.~Veretin,
Two-loop electroweak threshold corrections in the Standard Model,
Nucl. Phys. B \textbf{896} (2015), 19-51.
(\href{https://doi.org/10.1016/j.nuclphysb.2015.04.010}
{doi:10.1016/j.nuclphysb.2015.04.010})


\bibitem{Wells:2009kq}
J.~D.~Wells,
Lectures on Higgs Boson Physics in the Standard Model and Beyond,
[arXiv:0909.4541 [hep-ph]].

  
\bibitem{Jegerlehner:2014mua}
F.~Jegerlehner,
Higgs inflation and the cosmological constant,
Acta Phys. Polon. B \textbf{45} (2014) no.6, 1215.
(\href{https://doi.org/10.5506/APhysPolB.45.1215}
{doi:10.5506/APhysPolB.45.1215})
  

\bibitem{Masina:2013wja}
I.~Masina and M.~Quiros,
On the Veltman Condition, the Hierarchy Problem and High-Scale Supersymmetry,
Phys. Rev. D \textbf{88} (2013) 093003.
(\href{https://doi.org/10.1103/PhysRevD.88.093003}
{doi:10.1103/PhysRevD.88.093003})


\bibitem{Hamada:2012bp}
  Y.~Hamada, H.~Kawai and K.~y.~Oda,
  Bare Higgs mass at Planck scale,
  Phys.\ Rev.\ D {\bf 87} (2013) 053009;
   Erratum: [Phys.\ Rev.\ D {\bf 89} (2014) 059901].
(\href{https://doi.org/10.1103/PhysRevD.87.053009}
{doi:10.1103/PhysRevD.87.053009})


\bibitem{Sakharov:1967dj}
A.~D.~Sakharov,
Violation of CP Invariance, C asymmetry, and baryon asymmetry of the universe,
Pisma Zh. Eksp. Teor. Fiz. \textbf{5} (1967)  32-35;
JETP Lett. \textbf{5} (1967) 24-27.
(\href{https://doi.org/10.1070/PU1991v034n05ABEH002497}
{doi:10.1070/PU1991v034n05ABEH002497})


\bibitem{Isidori:2023pyp}
G.~Isidori, F.~Wilsch and D.~Wyler,
The Standard Model effective field theory at work,
[arXiv:2303.16922 [hep-ph]].


\end{thebibliography}
\end{document}